\documentclass[aps,pra,twocolumn,superscriptaddress]{revtex4-1}
\usepackage{amssymb,amsmath}
\usepackage{float}
\usepackage{stix}
\usepackage{graphicx}
\usepackage[dvipsnames]{xcolor}
\usepackage[colorlinks,linkcolor=blue,citecolor=blue,urlcolor=blue]{hyperref}
\usepackage{array}

\renewcommand{\[}{\left[}
\renewcommand{\]}{\right]}
\renewcommand{\(}{\left(}
\renewcommand{\)}{\right)}

\newcommand{\Langle}{\left\langle}
\newcommand{\Rangle}{\right\rangle}
\def\capo       {\right.\\ \left.}

\def\La         {\Langle}

\def\Ra         {\Rangle}

\def\wl         {\Big\{}
\def\wr         {\Big\}}
\newcommand{\bs}[1]{\boldsymbol{#1}}
\newcommand{\uu}[1]{\underline{#1}}
\newcommand{\oo}[1]{\overline{#1}}
\newcommand{\eq}[1]{Eq.\,\eqref{#1}}
\newcommand{\eqs}[1]{Eqs.\,\eqref{#1}}
\newcommand{\fig}[1]{Fig.\,\ref{#1}}
\newcommand{\h}[1]{\hat{#1}}
\newcommand{\ww}[1]{\widehat{#1}}
\renewcommand{\vec}[1]{\mathbf{#1}}

\def\bnu    {\uu{\nu}}
\def\bmu    {\uu{\mu}}
\def\bku    {\uu{\kappa}}
\def\blu    {\uu{\lambda}}
\def\bgs        {\underline{\sigma}}
\def\bgz        {\underline{\zeta}}

\def\bga        {\underline{\alpha}}
\def\bgb        {\underline{\beta}}

\def\bgl        {\underline\lambda}

\def\bgI        {\underline I}
\def\ga         {\alpha}
\def\gb         {\beta}
\def\gc         {\gamma}

\def\gd         {\delta}

\def\gl         {\lambda}

\def\gO         {\Omega}
\def\gr         {\rho}

\def\RR		{{\bf R}}

\newcommand{\xx}{\mathbf{x}}

\newcommand{\nn} {\nonumber}

\newcommand{\mE}{\mathcal{E}}

\begin{document}
\title{A  functional approach to the electronic and bosonic dynamics of many--body systems
  perturbed with an arbitrary strong electron--boson interaction}
\author{Andrea Marini} 
\affiliation{Istituto di Struttura della Materia of the National Research Council, via Salaria Km 29.3, I-00016 Monterotondo Stazione, Italy} 
\affiliation{European Theoretical Spectroscopy Facilities (ETSF} 
\affiliation{Division of Ultrafast Processes in Materials (FLASHit)}
\author{Yaroslav Pavlyukh}
\affiliation{Department of Physics and Research Center OPTIMAS,
  Technische Universit{\"a}t Kaiserslautern, P.O. Box 3049, 67653 Kaiserslautern, Germany} 
\affiliation{Institut f\"ur Physik, Martin-Luther-Universit\"at Halle-Wittenberg, 06120 Halle, Germany}
\date{\today}

\begin{abstract}
We present a formal derivation of the many--body perturbation theory for a system of
electrons and bosons subject to a nonlinear electron--boson coupling.  The interaction is
treated at an arbitrary high order of bosons scattered.  The considered Hamiltonian
includes the well--known linear coupling as a special limit. This is the case, for
example, of the Holstein and Fr\"{o}hlich Hamiltonians.  Indeed, whereas linear coupling
have been extensively studied, the scattering processes of electrons with multiple bosonic
quasiparticles are largely unexplored. We focus here on a self-consistent theory in terms
of dressed propagators and generalize the Hedin's equations using the Schwinger technique
of functional derivatives. The method leads to an exact derivation of the electronic and
bosonic self-energies, expressed in terms of a new family of vertex functions, high order
correlators and bosonic and electronic mean--field potentials. In the electronic case we
prove that the mean--field potential is the $n$th--order extension of the well--known
Debye--Waller potential. We also introduce a bosonic mean--field potential entirely
dictated by nonlinear electron--boson effects. The present scheme, treating electrons and
bosons on an equal footing, demonstrates the full symmetry of the problem. The vertex
functions are shown to have purely electronic and bosonic character as well as a mixed
electron--boson one. These four vertex functions are shown to satisfy a generalized
Bethe--Salpeter equation. Multi bosons response functions are also studied and explicit
expressions for the two and the three bosons case are given.
\end{abstract}
\maketitle
%
%
%
%
%
%
%
%
\section{Introduction}

Electron--boson\,(e--b) Hamiltonians are ubiquitous in particle, condensed matter physics
and optics: the fundamental electron--electron interaction is mediated by photons, which
are bosonic particles; lattice vibrations (phonons) play fundamental role in
superconductivity~\cite{giustino_electron-phonon_2017}; and collective excitations in
many--electron systems\,(plasmons) as well as bound electron--hole states\,(excitons) have
a bosonic nature.  Many examples of such a duality can also be found in strongly
correlated systems~\cite{gogolin_bosonization_1998}. The interaction between electrons and
bosons is typically treated \emph{linearly} in electronic density and bosonic
displacement~\cite{mahan_many-particle_2000}. The proportionality constant may have
different expressions depending on the microscopic details of the system.  

However, there are cases where nonlinear coupling is comparable in strength or even
dominate the first--order electron--boson interaction.

\paragraph{Electron--phonon coupling in quantum dots.} Very often the quadratic and
linear effects are inseparable, and the former can arise in, e.\,g., perturbative
elimination of the off--diagonal electron--phonon coupling in quantum dots.  For instance,
quadratic coupling of carriers in quantum dots to acoustic phonons modifies the
polarization decay and leads to exponential
dephasing~\cite{muljarov_dephasing_2004}. Linear coupling alone generates acoustic
satellites in the spectrum, but causes no Lorentzian
broadening~\cite{lundqvist_characteristic_1969,langreth_singularities_1970}.

\paragraph{Flexural phonons.} The balance between the first and the second--order effects can be
influenced by the symmetry. If a system possesses a mirror plane, the coupling to the
oscillations normal to this plane cannot be linear. This fact was noticed by Mariani and
von Oppen~\cite{mariani_flexural_2008} who demonstrated that \emph{flexural phonons}
couple \emph{quadratically} to the electron density. On the other hands, if the mirror
symmetry is broken by the presence of a substrate or by the gating, the coupling becomes
linear again~\cite{gunst_flexural-phonon_2017}. 

\paragraph{Holstein and Fr\"{o}hlich models.} The interplay
between the effects induced by different orders of the e--b interaction can have important
consequences in the Holstein model~\cite{holstein_studies_1959}.  This uses a simplified
form of the Fr\"{o}hlich Hamiltonian, where carriers couple to a branch of
 dispersionless optical phonons through a
momentum--independent coupling. In this case even small positive nonlinear interaction
reduces the effective coupling between the electrons and the lattice, suppressing
charge--density--wave correlations, and hardening the effective phonon
frequency~\cite{adolphs_going_2013,li_quasiparticle_2015}. These finding prompted further
theoretical investigations of the Holstein model with even more complicated double--well
electron--phonon interaction~\cite{adolphs_single-polaron_2014,adolphs_strongly_2014}
using a generalization of the momentum average approximation~\cite{berciu_greens_2006},
and of general form of interaction using the determinant quantum Monte Carlo
approach~\cite{li_effects_2015}.  Closely connected to these studies are recent
experiments emphasizing the role of nonlinear lattice dynamics as a mean for
control~\cite{forst_nonlinear_2011}, and as a basis for enhanced superconductivity in
MgB$_2$~\cite{liu_beyond_2001} and some cuprates~\cite{mankowsky_nonlinear_2014}. They
point toward large ionic displacement which is a prerequisite for the nonlinear
electron-phonon coupling.

\paragraph{Finite temperature effects.}
Another prominent example is the renormalization of electronic structures due to zero or
finite temperature phonons. As demonstrated by Heine, Allen and
Cardona\,(HAC)~\cite{allen_theory_1976,Cardona2005b} the linear and quadratic couplings
(in atomic displacement) are of the same order in the electron--ion interaction
potential. Moreover they need to be considered on an equal footing in order for the system
to respect the system translational invariance. The effect of the second--order correction
is quite large in carbon materials and can lead to a substantial band gap
modification~\cite{giustino_electron-phonon_2010,cannuccia_effect_2011,ponce_temperature_2015}.

\paragraph{Anharmonic effects.}
Some recent works have also demonstrated that, potentially, even simple systems like
diamond~\cite{Antonius2015,Monserrat2013a} or palladium~\cite{errea_first-principles_2013}
show remarkable nonlinear effects. However, at the moment, these anharmonic effects can be
treated only by using an adiabatic approach based on finite displacements of the atoms.
This approach ignores dynamic effects that, however, have been
shown to be relevant in the linear coupling case~\cite{cannuccia_effect_2011} and,
therefore, cannot be neglected, {\em a priori} in the case of nonlinear coupling.

\paragraph{Existing theoretical approaches.}
Nonlinear electron--boson models have been treated theoretically by essentially stretching
methods developed for pure electronic case or linear coupling scenario: quantum Monte
Carlo~\cite{li_quasiparticle_2015}, the average momentum
approximation~\cite{berciu_greens_2006}, and the cumulant
expansion~\cite{muljarov_dephasing_2004}. Since only electronic spectrum was of interest,
they rely on diagrammatic methods, without systematically exploring the
\emph{renormalization of phononic properties} due to electrons. However, as has been shown
in the linear case using perturbative expansions of both electron and phonon propagators,
electrons typically overscreen bare phonon frequencies leading to the conclusion that
renormalized phonon frequencies must be fitted to
experiments~\cite{van_leeuwen_first-principles_2004}. Thus, Marini \emph{et
  al.}~\cite{marini_many-body_2015} have recently extended many-body perturbation
theory\,(MBPT) for electron--phonon interaction including quadratic terms and using
Density Functional Theory~\cite{R.M.Dreizler1990} as a starting point. This is a
remarkable achievement since even \emph{ab initio} determination of momentum-dependent
electron-phonon linear coupling function is a nontrivial task~\cite{verdi_frohlich_2015}.
The Born--Oppenheimer approximation is commonly used as a starting point. However,
  the seminal works of Abedi \emph{et al.}~\cite{PhysRevLett.105.123002} and Requist
  \emph{et al.}~\cite{PhysRevLett.117.193001} on the exact factorization of the fermionic
  and bosonic wave--function show that alternative paths beyond the Born--Oppenheimer
  approximation are possible.

\paragraph{Diagrammatic perturbation theory}
Nonexistence of the Wick theorem for bosons~\cite{abrikosov_methods_1975}, which is a
consequence of the fact that averages of the normal product of bosonic operators are
non-zero, makes it difficult to develop a diagrammatic perturbation
theory~\cite{pavlyukh_pade_2017}. To circumvent this difficulty, systems above the
Bose-condensation temperature are implicitly assumed~\cite{van_leeuwen_wick_2012}. Method
of functional derivatives is a complementary method~\cite{strinati_application_1988}. In
contrast to diagrammatic constructions based on the series expansions of the evolution
operator on a contour, it yields functional relations between the dressed
propagators. They do not rely on the Wick theorem. In the seminal works of
L.\,Hedin~\cite{Hedin19701} and R.\,van\,Leeuwen\cite{van_leeuwen_first-principles_2004},
the Schwinger technique of functional derivatives is used to derive the linear
electron--boson coupling and no Debye--Waller\,(DW) potential is found. This is in
stringent disagreement with the HAC theory where this potential naturally appears.  On the
other hand any diagrammatic approach predicts the existence of the DW potential, as, for
example in Ref.~\onlinecite{marini_many-body_2015}.  It is therefore desirable to formulate
\emph{self-consistent} (\emph{sc}) MBPT for electron--boson system with nonlinear
coupling, i.\,e., in terms of the \emph{dressed} propagators, in a functional derivative
approach.

\paragraph{Out--of--equilibrium scenarios}
Our further motivation for this work is experimental
feasibility to generate \emph{coherent}
phonons~\cite{papalazarou_coherent_2012,yang_ultrafast_2014} and
plasmons~\cite{huber_femtosecond_2005,lemell_real-time_2015,schuler_electron_2016}. For
such scenarios the notion of transient spectral properties is of special
interest~\cite{basov_electrodynamics_2011,moskalenko_attosecond_2012,perfetto_ultrafast_2018}. A
powerful method to deal with time-dependent processes is the non-equilibrium Green's
function (NEGF) approach~\cite{stefanucci_nonequilibrium_2013}. The method relies on
solving the Kadanoff-Baym equations (KBE) of motion for the Green's functions (GFs) on the
Keldysh time
contour~\cite{dahlen_solving_2007,myohanen_kadanoff-baym_2009,von_friesen_successes_2009,
  schuler_time-dependent_2016,schlunzen_dynamics_2016}. To the best of our knowledge, for
systems with nonlinear coupling such theory is not available.

{\em Manuscript organization}. Our manuscript is organized as follows: In Sec.~\ref{sec2}
we introduce the Hamiltonian and its properties. Given the Hamiltonian, in Sec.~\ref{sec:time:dep},
we derive the corresponding equation of motion for the bosonic and electronic operators.
The equation of motion are analyzed in terms of functional derivatives in
Sec.~\ref{sec:func:dir}. The Green's functions are introduced in Sec.~\ref{sec_GFs}.

We first discuss the electronic case whose self-energy is derived {\em exactly} to all
orders in the electron--boson interaction, in Sec.~\ref{sec_fermi_SE}. We derive the form
of a generalized Debye--Waller potential in Sec.~\ref{sec_DW} which, in turns, define the
remaining nonlocal and time--dependent mass operator, Sec.~\ref{sec_mass_operator}.

The bosonic subsystem is, then, split in single--boson and multi--boson case in analogy
with the electronic case. In Sec.~\ref{sec_single_boson} we introduce the bosonic
self-energy that we split {\em exactly} in a mass operator, Sec.~\ref{sec_boson_sigma},
and a mean--field potential, Sec.~\ref{sec_boson_mean_field}.  The exact bosonic mass
operator is rewritten in terms of four generalized vertex functions whose coupled equation
of motion is derived in Sec.~\ref{sec_boson_vertexes}.

The presented exact formulation is illustrated by
  the derivations of the lowest order approximations for the
electronic\,(Sec.~\ref{sec_Fan}) and bosonic\,(Sec.~\ref{sec_bose_Fan}) self-energies.

The last part of the work is devoted to the electronic and bosonic response
functions\,(Sec.~\ref{sec:responses}). We derive a Bethe--Salpeter like equation for the
electronic response in Sec.~\ref{sec:electronic_response}. In Sec.\ref{sec:resp_bosons} we
discuss the bosonic case by showing how to reduce the general bosonic dynamics to
diagonal number conserving response functions. Then, the cases of two and three bosons
are studied, respectively, in Sec.~\ref{sec:two_bosons} and Sec.~\ref{sec:D33}.

\begin{figure}[]
  {\centering
  \includegraphics[width=\columnwidth]{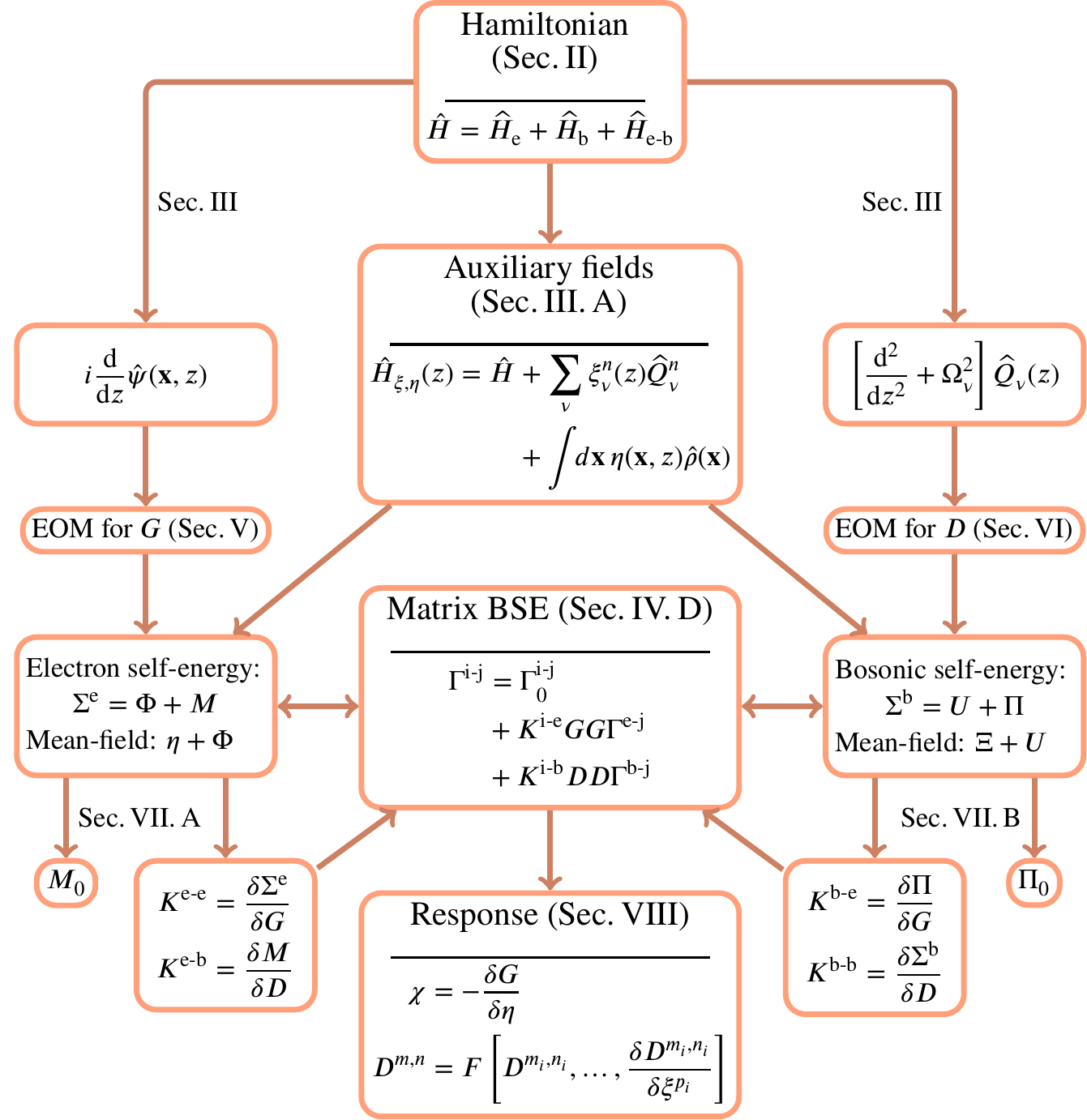}}
    \caption{Logical structure of the work.\label{fig:toc}}
\end{figure}

Finally, in Appendix~\ref{mean_field} we motivate our treatment of electron--electron
correlation, in Appendix~\ref{e-p-connect} we formally connect the formalism to the
electron--phonon problem. In Appendix~\ref{definitions} we finally list some key
mathematical quantities and approximations used throughout the whole manuscript.
Logical flow of the whole work is depicted on Fig.~\ref{fig:toc}.
%
%
%
%
%
%
%
%
\section{Notation and Hamiltonian}\label{sec2}
We start from the generic form of the total Hamiltonian of the system that we assume to be
composed by fermions and bosons with a nonlinear interaction
\begin{align}
\ww{H}= \ww{H}_\text{e}+\ww{H}_\text{b}+\ww{H}_\text{e--b}.
\label{eq:1.0}
\end{align}
The unperturbed part of $\ww{H}$ is $\ww{H}_\text{e}+\ww{H}_\text{b}$ and can be rewritten
in terms of corresponding energies ($\mE_i$ is the energy of the electronic state $i$,
$\Omega_\nu$ is the energy of the bosonic mode $\nu$) and eigenstates obeying fermionic,
bosonic statistics, respectively:
\begin{subequations}
  \label{eq:1.1}
\begin{gather}
 \ww{H}_\text{e}=\sum_i \mE_i \h{c}^{\dagger}_i \h{c}_i,\label{eq:H:e}\\
 \ww{H}_\text{b}=\frac12\sum_\nu \Omega_\nu\(\ww{P}_\nu^2+\ww{Q}_\nu^2\).
\end{gather}
\end{subequations}
In general, the partitioning of a physical Hamiltonian in the form of \eq{eq:1.0} is an
highly nontrivial problem~\cite{PhysRevLett.105.123002,PhysRevLett.117.193001}.  In the
present context, we are interested in the nonlinear e--b coupling and, to keep the
formulation simple, we assume that such a partition does exist and that the electronic
correlation can be approximatively described with a mean--field potential that
renormalizes the free electrons and bosons. This is a common practice, for example, in the
DFT approach to electrons and phonons. The DFT mean--field potential is defined in
Appendix~\ref{mean_field}.

In Eq.(\ref{eq:1.1}b) we have introduced the operators for the bosonic coordinates,
$\ww{Q}_{\nu}$, and momenta, $\ww{P}_{\nu}$.  The fermions are described by the
corresponding creation\,($\h{c}^{\dagger}_i$) and annihilation\,($\h{c}_i$)
operators. These are used to expand the electronic field operator $\hat\psi\(\xx\)=\sum_i
\phi_i\(\xx\) \h{c}_i$, with $\phi_i\(\xx\)$ eigenfunctions of the electronic Hamiltonian
in the first quantization (denoted as $h_\text{e}(\xx)$).

$\mE_i$ and $\Omega_\nu$ are the independent electrons and bosons energies. They are
assumed to incorporate the mean--field potentials embodied in
$\ww{H}_\text{e}+\ww{H}_\text{b}$. $\h{Q}_{\nu}$ and $\h{P}_{\nu}$ are expressed in the
standard way in terms of the creation\,($\h{b}^{\dagger}_\nu$) and
annihilation\,($\h{b}_\nu$) operators:
\begin{subequations}
  \label{eq:1.1p}
\begin{gather}
\ww{Q}_\nu = \frac{1}{\sqrt{2}}\left(\hat{b}^\dagger_\nu + \hat{b}_\nu\right),\\
\ww{P}_\nu = \frac{i}{\sqrt{2}}\left(\hat{b}^\dagger_\nu - \hat{b}_\nu\right).
\end{gather}
\end{subequations}
The electron--boson interaction is taken to have the general form:
\begin{align}
\ww{H}_\mathrm{e-b}=
\sum_{n,\,\bnu} \int\!d\xx\,\hat \psi^\dagger\(\xx\) V^{n}_{\bnu}\(\xx\) \hat\psi\(\xx\)\ww{Q}^n_{\bnu},
\label{eq:1.2}
\end{align}
with 
\begin{subequations}
  \label{eq:1.3}
\begin{gather}
 \ww{Q}^n_{\bnu}=\prod_{i=1}^n \ww{Q}_{\nu_i},\\
 V^n_{\bnu}\(\xx\)=\frac{1}{n!}\(\prod_{i=1}^n \partial_{\nu_i}\)_{eq} V_\text{e--b}\(\xx\).
 \label{eq:Vn}
\end{gather}
\end{subequations}
Here, $V_\text{e--b}\(\xx\)$ is a generic potential that dictates the electron--boson
interaction.  The connection to the electron--phonon problem is given in the
Appendix\,\ref{e-p-connect}.  Eq.\eqref{eq:Vn} makes it clear that $V^n_{\bnu}\(\xx\)$ is
a \emph{symmetric tensor} with respect to indices $\bnu$.  The differentiation is
performed with respect to the bosonic coordinates evaluated at the equilibrium point.  The
physical form of the potential depends on the specific problem.  Therefore the equilibrium
coordinates are specific to the kind of physics the bosons are describing. In the case of
phonons $\(\prod_{i=1}^n \partial_{\nu_i}\)_{eq}$ is evaluated at the equilibrium atomic
configuration, as defined in Appendix\,\ref{e-p-connect}.

Averaging the total Hamiltonian, \eq{eq:1.0}, with respect to electronic coordinates leads
to the effective \emph{anharmonic} bosonic Hamiltonian.  Solving such a model leads to,
among other effects, the prediction of the temperature dependence of the averaged
displacement. While interesting and well-discussed problem on its own, we will not
consider this effect here assuming that for each given temperature an Hamiltonian of the
type defined by \eq{eq:1.0} can be derived such that
\begin{equation}
  \Langle \ww{Q}_\nu \Rangle=0.
  \label{eq:Q0}
\end{equation}
In contrast, as will be shown using our diagrammatic approach, other correlators of the
position operator will be modified by electron-boson interaction in nontrivial way.

\eq{eq:1.3} highlights an important and crucial aspect of the notation. The symbol $\bnu$
represents a generic vector of bosonic indices of dimension $n$, which is indicated as a
superscript and should not be confused with power.  Therefore we consider the most general
case where the $n$th--order e--b interaction is a nonlocal function of $n$ bosonic
coordinates.

For convenience we also introduce the electronic operator 
\begin{align}
\hat \gc^{n}_{\bnu}\equiv \int\!d\xx\,\hat \psi^\dagger\(\xx\)V^{n}_{\bnu}\(\xx\) \hat \psi\(\xx\),
\end{align}
such that $ \ww{H}_\mathrm{e-b}$ can be written as
\begin{align}
  \ww{H}_\mathrm{e-b} = \sum_{n,\,\bnu} \hat{\gc}^{n}_{\bnu} \ww{Q}_{\bnu}^n.
  \label{eq:h:el-bos}
\end{align}
Having introduced the general electron-boson Hamiltonian~\eqref{eq:1.0} and
specified its ingredients, our goal now is to obtain a self-consistent set of equations
that relate well-defined objects such as electron and boson propagators. To this end, we
generalize the Schwinger's method of functional derivatives~\cite{martin_theory_1959},
which allows to express more complicated correlators that appear in their equations of
motion (the Martin-Schwinger hierarchy) in terms of functional derivatives.
%
%
%
%
%
%
%
\section{The Equation of Motion for the electronic and bosonic operators}\label{sec3}
\subsection{Time Dependence\label{sec:time:dep}}
For our purpose we define operators in the Heisenberg picture (indicated here by
  the $H$ subscript) with time--arguments running on the Keldysh contour ($z\in{\mathcal
  C}$):
\begin{align}
\ww{\mathcal O}_H\(z\)\equiv \ww{\mathcal U}\(z_0,z\)\ww{\mathcal O}\,\ww{\mathcal U}\(z,z_0\),
\end{align}
where $z_0$ is arbitrary initial time and $\hat{\mathcal U}\(z,z_0\)$ is the
time-evolution operator from the initial time $z_0$ to $z$. In this picture, the
  operators are explicitly time--dependent, whereas wave-functions not. This allows to make
  a connection with the many-body perturbation theory, which relies on the time-evolution
  on the contour and on the Wick theorem. In what follows, the picture in which operators
  are given is not explicitly indicated when it can be inferred from the corresponding
  arguments.

The electronic, bosonic operators satisfy standard anticommutation (denoted with $+$),
commutation (denoted with $-$) rules, respectively:
\begin{subequations}
  \label{eq:comm}
\begin{align}
\[\hat \psi\(\xx_1\),\hat   \psi^\dagger\(\xx_2\)\]_{+}&=\delta\(\xx_1-\xx_2\),\\ 
\[\ww{Q}_\mu\,,\ww{P}_{\nu}\]_{-}&=i\delta_{\mu\nu}.
\end{align}
\end{subequations}
We now introduce a short-hand notation $\(\xx_i,z_i\)\equiv i$ so that $\hat
\psi\(1\)\equiv\hat\psi\(\xx_1,z_1\)$.  The Heisenberg equations of motion\,(EOM) for
$\hat\psi$, $\ww{Q}$ and $\ww{P}$ follow by applying Eqs.\eqref{eq:comm} to evaluate commutators with the full Hamiltonian $\ww{H}$:
\begin{subequations}
  \label{eq:deriv}
\begin{gather}
 i\frac{d}{dz_1}\hat \psi\(1\)=\[h_\text{e}\(1\)+\sum_{n,\,\bnu} V^{n}_{\bnu}\(\xx_1\)\ww{Q}^n_{\bnu}\(z_1\) \]\hat \psi\(1\),\label{eq:dpsi}\\
  \frac{d}{dz_1}\ww{Q}_\nu\(z_1\)=\Omega_\nu \ww{P}_\nu\(z_1\),\label{eq:dQ}\\
  \frac{d}{dz_1}\ww{P}_\nu\(z_1\)=-\Omega_\nu \ww{Q}_\nu\(z_1\)
  -\sum_{m,\,\bmu} m\, \h{\gamma}^{m}_{\bmu\oplus\nu}\(z_1\)\ww{Q}^{m-1}_{\bmu}\(z_1\).\label{eq:dP}
\end{gather}
\end{subequations}
In \eq{eq:dP}, the combinatorial prefactor $m$ follows from the fact that $\h{\gamma}^{m}$
also is a symmetric tensor of rank $m$. This equation is formally demonstrated in
Appendix\,\ref{app_eq_dP}.

In Eq.\,(\ref{eq:dP}) we have introduced a general definition for a multi--dimensional
operator whose index is a composition of two subgroups of indexes. In the case of
$\h{\gamma}^{m}_{\bmu\oplus\nu}$, the vector of indices $\bmu$ has $m-1$ components, and
$\big(\bmu\oplus\nu\big)=\(\mu_1,\dots\mu_{m-1},\nu\)$ is correctly $m$ dimensional.  By
combining the last two of \eqs{eq:deriv} we obtain a second--order differential equation
for the displacement operator~\footnote{In the case of out--of--equilibrium dynamics the
  one boson displacement operator
fulfills an harmonic oscillator equation driven by the time-dependent electron
density~\cite{schuler_time-dependent_2016}} with a source term:
\begin{align}
  \left[\frac{d^2}{dz_1^2}+\Omega_\nu^2\right]\ww{Q}_\nu\(z_1\)
  =-\Omega_\nu\-\sum_{m,\,\bmu} m\, \hat\gamma^{m}_{\bmu\oplus\nu}\(z_1\)\ww{Q}^{m-1}_{\bmu}\(z_1\).
  \label{eom:Q}
\end{align}
More compicated operators appearing on the right hand side of
Eqs.~(\ref{eq:deriv},\ref{eom:Q}) can be expressed using the method of functional
derivatives.
\subsection{Functional derivatives\label{sec:func:dir}}
In order to introduce the functional derivatives approach we couple the Hamiltonian to
\emph{time-dependent auxiliary fields} $\xi^{n}_{\bnu}\(z\)$ and $\eta\(\xx,z\)$
\begin{align}
  \hat{H}_{\xi,\eta}\(z\)=\hat{H}
  +\sum_{n,\,\bnu}\xi^{n}_{\bnu}\(z\)\ww{Q}_{\bnu}^n\(z\) + \int\!\!d\xx\, \eta\(\xx,z\)\hat{\gr}\(\xx,z\),
\label{eq:sources}
\end{align}
where a superscript in $\xi^{n}_{\bnu}\(z\)$ indicates that $\bnu$ is an $n$-dimensional
vector of indices. We introduced the electron density operator
$\hat{\rho}(1)=\hat\psi^\dagger(1)\hat\psi(1)$. 

Consider now the time-evolution in the
presence of these external fields. The corresponding time-evolution operator is denoted as
$\ww{\mathcal U}_{\xi,\eta}\(z_0,z\)$. Now in the definition of the average operator  
\begin{align}
  \Langle\hat{\mathcal O}_{\xi,\eta}\(z\)\Rangle_{\xi,\eta}
  =\frac{\mathrm{Tr}\Big\{{\mathcal T}
    \exp\big[-i\int_{\mathcal C}d\bar{z}\ww{H}_{\xi,\eta}(\bar{z})\big]\,
    \hat{\mathcal O}_{\xi,\eta}\(z\)\Big\}}{\mathrm{Tr}\Big\{{\mathcal T}
    \exp\big[-i\int_{\mathcal C}d\bar{z}\ww{H}_{\xi,\eta}\(\bar{z}\)\big]\Big\}},
  \label{diff_form_xi}
\end{align}
the $\xi$ and $\eta$ functions occur twice signaling that both: the operator $\hat{O}$ in the
Heisenberg picture $\hat{\mathcal O}_{\xi,\eta}\(z\)=\ww{\mathcal
  U}_{\xi,\eta}\(z_0,z\)\hat{\mathcal O}\ww{\mathcal U}_{\xi,\eta}\(z,z_0\)$ and the
density matrix are defined with respect to the perturbed Hamiltonian.  Starting from this
form various functional derivatives can be computed. We write $\Langle\ldots
\Rangle$ for $\Langle\ldots\Rangle_{\xi,\eta}$ where it does not lead to
ambiguities.
\begin{widetext}
Let us consider the case of a generic, contour--ordered product of operators:
$\prod_i\hat{\mathcal O}^{\(i\)}\(z_i\)$. Constituent operators depend, in general, on
different times $z_i$ and are distinguished by the subscript $(i)$. By the formal
differentiation, one can prove that
\begin{align}
  i\left.\frac{\delta}{\delta\xi^{n}_{\bmu}\(\bar{z}\)}\Big\langle{\mathcal T}
  \Big\{\prod_i\hat{\mathcal O}^{\(i\)}_{\xi,\,\eta}\(z_i\)\Big\}\Big\rangle\right|_{\xi^{n}_{\bmu}=0,\,\eta=0}=
  \Big\langle{\mathcal T} \Big\{\prod_i\hat{\mathcal O}^{\(i\)}\(z_i\)\,\ww{Q}^n_{\bmu}\(\bar{z}\) \Big\}\Big\rangle
  -\Big\langle{\mathcal T} \Big\{\prod_i\hat{\mathcal O}^{\(i\)}\(z_i\) \Big\}
  \Big\rangle \Big\langle\ww{Q}^n_{\bmu}\(\bar{z}\)\Big\rangle,
  \label{id:diff}
\end{align}
where $\mathcal{T}$ denotes the contour ordering operator.  The second term in
\eq{id:diff} stems from the variation of denominator, i.\,e., it assures correct
normalization. In general, this identity can contain side by side electronic and bosonic
operators and also operators with equal time arguments. For the latter, the standard
definition of $\mathcal T$ needs to be amended with a rule that equal-time operators do
not change their relative order upon contour-ordering. For mixed operators, only the
permutations of the electronic ones induce a
sign--change~\cite{stefanucci_nonequilibrium_2013}.

A similar expression holds for the derivative with respect to $\eta$:
\begin{align}
  i\left.\frac{\delta}{\delta\eta\(1\)}\Big\langle{\mathcal T}
  \Big\{\prod_i\hat{\mathcal O}^{\(i\)}_{\xi,\,\eta}\(z_i\)\Big\}\Big\rangle\right|_{\xi^{n}_{\bmu}=0,\,\eta=0}=
   \Big\langle{\mathcal T}\Big\{\prod_i\hat{\mathcal O}^{\(i\)}\(z_i\)\,\gr\(1\)\Big\}\Big\rangle-
   \Big\langle{\mathcal T}\Big\{\prod_i\hat{\mathcal O}^{\(i\)}\(z_i\)\Big\}\Big\rangle \Big\langle\gr\(1\)\Big\rangle.
  \label{id:diff_prime}
\end{align}
\end{widetext}
Here and in the following we always assume that the limit of zero auxiliary fields is
taken after variations. In practice, however, this means that during derivations all
Green's functions are formally dependent on the auxiliary fields. This will be evident
from the form of the electronic and bosonic Dyson equations with mean--fields that include
the auxiliary fields.
%
%
%
%
%
%
%
%
\section{Green's function and diagrammatic notation}\label{sec_GFs}
We use the standard definitions of the electronic Green's function (GF) on the Keldysh contour:
\begin{align}
  G\(1,2\)=-i\Langle\mathcal{T}\{\hat \psi\(1\)\hat\psi^\dagger\(2\)\}\Rangle,
\end{align}
where $\langle\ldots\rangle$ is the trace evaluated with the exact density matrix.

The bosonic propagators on the Keldysh contour extend the definition of the electronic case
\begin{align}
  D^{m,n}_{\bmu,\bnu}\(z_1,z_2\)=
  -i\Langle\mathcal{T}\wl\Delta\ww{Q}^m_{\bmu}\(z_1\)\Delta\ww{Q}^n_{\bnu}\(z_2\)\wr\Rangle,
\end{align}
where $\Delta \hat {\mathcal O}\equiv\hat {\mathcal O}-\langle\hat{\mathcal O}\rangle$ is
the fluctuation operator. In the case $m=n=1$ the standard bosonic propagator is recovered
\begin{align}
 D_{\mu,\nu}\(z_1,z_2\)=D^{1,1}_{\mu,\nu}\(z_1,z_2\).
\end{align}

\begin{figure}[t!]
  {\centering
  \includegraphics[width=\columnwidth]{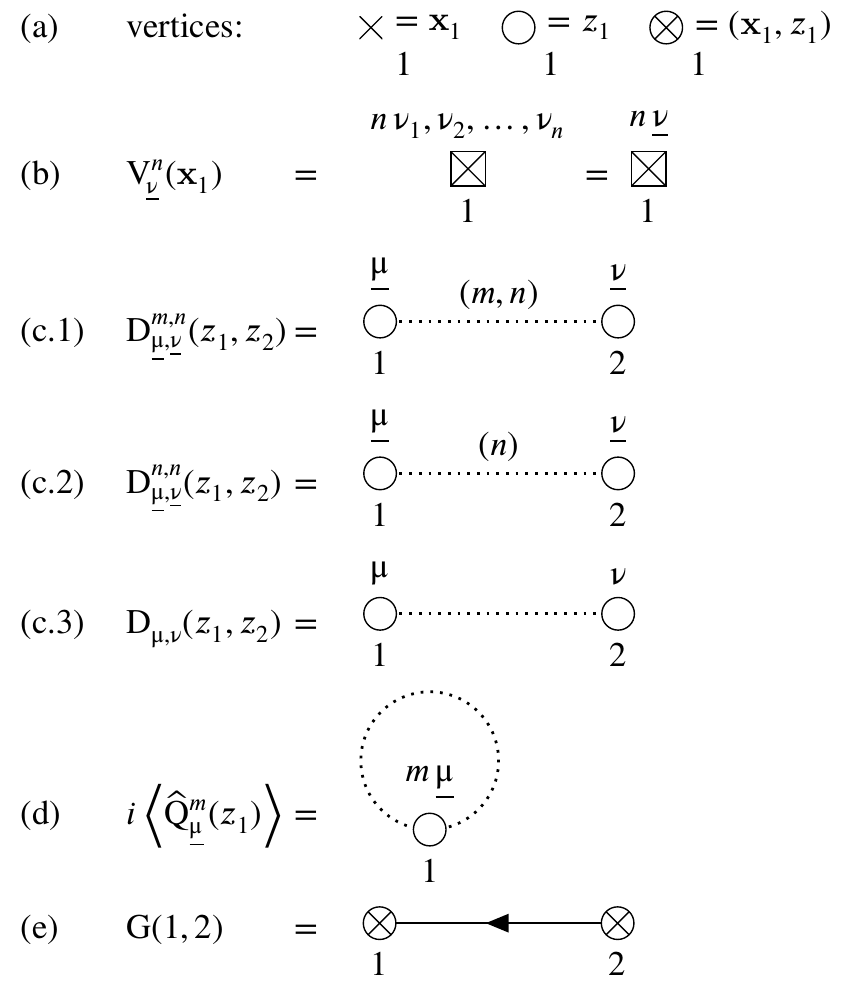}}
    \caption{Definition of the diagrammatic elements used in this work. (a) $\bigcirc$ and
      $\times$ represent a generic time and position point respectively.  These two
      symbols can be combined to indicate a time and position vertex $\oplus_1$
      equivalent to $1=\(\xx_1,z_1\)$. (b) Finally a box around a spatial point represents
      the scattering integral $V^{n}_{\bnu}\(\xx\)$ with two fermionic and $n$ bosonic
      dangling lines. (c) Bosonic propagators can be represented in three different forms
      depending on their order. (d) Expectation value of the bosonic coordinates expressed
      in terms of a bosonic propagator. (e) Electronic Green's function.\label{fig:1}}
\end{figure}

Thanks to \eq{diff_form_xi}, we can rewrite $D^{m,n}$ as
\begin{multline}
iD^{m,n}_{\bmu,\bnu}\(z_1,z_2\)=  \Langle {\mathcal T}\ww{Q}^m_{\bmu}\(z_1\)\ww{Q}^n_{\bnu}\(z_2\)\Rangle\\
-\Langle \ww{Q}^{m}_{\bmu}\(z_1\)\Rangle\Langle \ww{Q}^{n}_{\bnu}\(z_2\)\Rangle
=  i\frac{\delta \Langle \ww{Q}_{\bmu}^m\(z_1\)\Rangle}{\delta\xi_{\bnu}^{n}\(z_2\)}.
 \label{eq:D_as_dQ_dxi}
\end{multline}
This equation can be further generalized to
\begin{multline}
  D^{m,n}_{\bmu,\bnu}\(z_1,z_2\)= i\frac{\delta}{\delta\xi_{\bku}^{k}\(z_1\)}D^{m-k,n}_{\blu,\bnu}\(z_1,z_2\)\\
  +\Langle \ww{Q}_{\bku}^k\(z_1\)\Rangle D^{m-k,n}_{\blu,\bnu}\(z_1,z_2\)\\
  +\Langle \ww{Q}^{m-k}_{\blu}\(z_1\)\Rangle D^{k,n}_{\bku,\bnu}\(z_1,z_2\),
 \label{eq:D_as_dD}
\end{multline}
for $k<m$ and $\bmu=\bku\oplus\blu$. \eq{eq:D_as_dD} is proved in
Appendix\,\ref{app_D_as_dD_proof}. The last two terms represent a contraction of symmetric
tensors of ranks $m-k$ and $k$ yielding a symmetric tensor of rank $m$ (with respect to
the first argument). We will make an extensive use of these differential form of $D^{m,n}$
as well as of the representation in terms of Feynman diagrams. We introduce {\em ad hoc}
graphical objects to easily represent the multi-fold aspects of the nonlinear e--b
interaction; in \fig{fig:1} all ingredients of the diagrammatic representation are showed.

In general, the selection of $k$ bosonic operators out of $m$, that appear on the
r.h.s. of \eq{eq:D_as_dD}, can be performed in $\binom{m}{k}$ ways. These corresponds to
all the possible choices of $k$ elements out of $m$. However \eq{eq:D_as_dD} is {\em
  exact} for any choice of the $\bku$ elements. Therefore no combinatorial prefactor is
needed whenever \eq{eq:D_as_dD} is used.

By using \eq{eq:Q0} we can write
\begin{align}
  \Langle \ww{Q}^{m}_{\bmu\oplus\nu}\(z_1\)\Rangle = iD^{m-1,1}_{\bmu,\nu}\(z_1,z_1^+\).
 \label{eq:Q_average}
\end{align}
We use here $z_1^+=z_1+0^+$.  It is important to note, here, that in the limit
$\xi^{n}_{\bmu}=0,\,\eta=0$ we have that $\Langle
\ww{Q}^{m}_{\bmu\oplus\nu}\(z_1\)\Rangle$ is
constant because of the time--translation invariance. However during the derivation the
time-dependence is induced by the auxiliary fields.

The EOM for bosonic displacement operators~\eqref{eom:Q} leads us to consider a
specific case of $D^{m-1,1}$, which can be reduced to simpler propagators by the
application of \eq{eq:D_as_dD} with $k=m-2$:
\begin{multline}
 D^{m-1,1}_{\bmu,\nu}\(z_1,z_2\)= i\frac{\delta}{\delta\xi_{\bku}^{m-2}\(z_1\)}D_{\gl,\nu}\(z_1,z_2\)\\ 
  +\Langle \ww{Q}_{\bku}^{m-2}\(z_1\)\Rangle D_{\gl,\nu}\(z_1,z_2\),
 \label{eq:D_mm1_1_as_dD}
\end{multline}
where we used the fact that $\langle \ww{Q}_{\lambda}\rangle$ is zero in the limit of
vanishing auxiliary fields and $\bmu=\bku\oplus\gl$.
%
%
%
%
%
%
\section{Electron dynamics}~\label{sec_fermi_SE}
The EOM for $G$ is obtained with the help of EOMs for the constituent operators and using
the relation $\frac{d}{dz_1}\theta\(z_1-z_2\)=\delta\(z_1-z_2\)$. Thus, we have
\begin{multline}
  \[i\frac{\partial}{\partial z_1}-h_\text{e}\(1\)-\eta\(1\)\]G\(1,2\)=\delta\(1,2\)\\
  -i\sum_{n,\,\bnu}V_{\bnu}^{n}\(\xx_1\) \Langle\mathcal{T} \wl \hat
  \psi\(1\)\ww{Q}_{\bnu}^n\(z_1\)\hat \psi^\dagger\(2\) \wr \Rangle.\label{eom:G}
\end{multline}
Using \eq{id:diff}, the correlator on r.h.s. of Eq.~\eqref{eom:G} can be expressed
  as the functional derivative
\begin{multline}
  -i\Langle\mathcal{T} \hat \psi\(1\)\ww{Q}_{\bnu}^n\(z_1\)\hat \psi^\dagger(2)\Rangle\\
  =\[i\frac{\delta}{\delta\xi^{n}_{\bnu}\(z_1\)}
  +\Langle \ww{Q}_{\bnu}^n\(z_1\) \Rangle\] G\(1,2\).\label{eq:var1}
\end{multline}

Our goal is to rewrite Eq.~\eqref{eom:G} in the form of a Dyson equation, which involves a
dressed mean--field potential $\Phi$ and correlated mass operator $M$:
\begin{multline}
  \[i\frac{\partial}{\partial z_1}-h_\text{e}\(1\)-\eta\(1\)-\Phi\(1\)\]G(1,2)=\delta(1,2)\\
  +\int\!d3\,M\(1,3\)G\(3,2\).
\end{multline}
The potential $\Phi$ follows from the second term on the r.h.s. of Eq.~\eqref{eq:var1}
\begin{align}
 \Phi\(1\)=\sum_{n,\,\bnu}V_{\bnu}^{n}\(\xx_1\)\Langle \ww{Q}_{\bnu}^n\(z_1\)\Rangle.
 \label{eq:v_e}
\end{align}
\begin{figure}[t!]
{\centering
\includegraphics[]{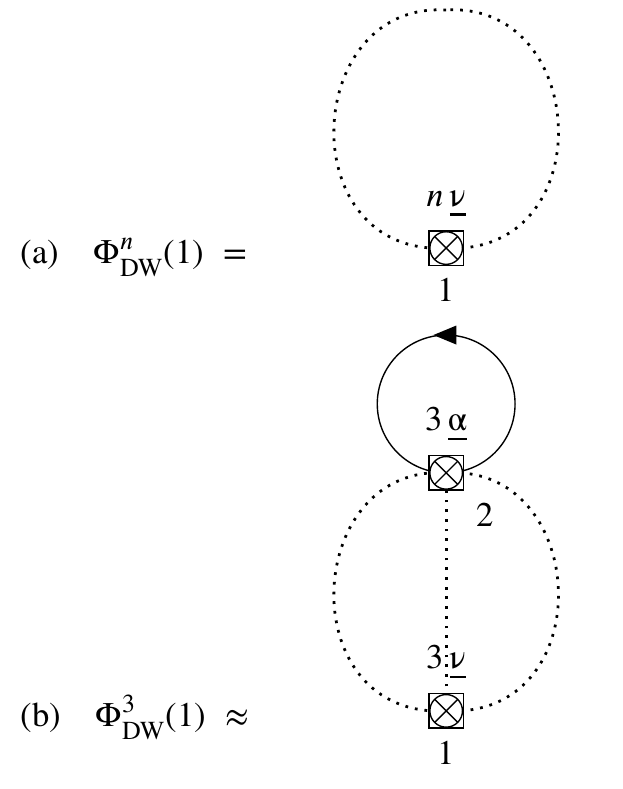}}
\caption{(a) Diagrammatic form of the $n$th--order DW potential. (b) Perturbative
  expansions in term of \emph{bare} bosonic propagators (they are denoted as dashed lines)
  lead to complicated diagrams. The inclusion of nonlinear e-b interaction leads to
  non-vanishing odd-order terms that are zero in the linear interaction
  case.\label{fig:DW}}
\end{figure}
The mass operator is implicitly written as
\begin{align}
  \int\!d3\,M\(1,3\)G\(3,2\)&
  = i\sum_{n,\,\bnu} V_{\bnu}^{n}\(\xx_1\)\frac{\delta}{\delta\xi_{\bnu}^{n}\(z_1\)}G\(1,2\).
  \label{eq:mass_e}
\end{align}
The potential $\Phi$ and the mass operator $M$ can be conveniently combined in the
electronic self--energy operator $\Sigma^\text{e}$:
\begin{align}
  \Sigma^\text{e}\(1,2\)= \Phi\(1\)\delta(1,2)+M\(1,2\).
  \label{eq:sigma_e}
\end{align}
%
%
%
%
%
\subsection{The $n$th--order Debye--Waller potential}~\label{sec_DW}
In order to rewrite $\Phi$ in terms of the bosonic Green's function, we apply
\eq{eq:Q_average} to \eq{eq:v_e}. It follows that we can introduce a $n$th--order bosonic
mean field, $\Phi^{n}_\text{DW}\(1\)$, defined as:
\begin{align}
  \Phi^{n}_\text{DW}\(1\)=i\sum_{\bnu}V_{\bnu}^{n}\(\xx_1\)D^{n-1,1}_{\bmu,\nu_n}\(z_1,z_1^+\),
  \label{eq:v_dw_2}
\end{align}
with $\bnu=\bmu\oplus \nu_n$. $\Phi^{n}_\text{DW}$ is showed in diagrammatic form in
Fig.\,\ref{fig:DW}(a) in the general case.

\eq{eq:v_dw_2} provides a generalization of the Debye--Waller\,(DW) potential to arbitrary
orders. The expression of this potential is well--known in the electron--phonon case only
when $n=2$, and it has been derived only by using a diagrammatic approach. In the present
case, it naturally appears as the mean--field electronic potential induced by the
nonlinear electron--boson interaction:
\begin{align}
 \Phi^{2}_\text{DW}\(1\)=i\sum_{\nu_1,\nu_2}V_{\nu_1,\nu_2}^{2}\(\xx_1\)D_{\nu_1,\nu_2}\(z_1,z_1^+\).
\end{align}
The DW potential has a long history in the electron--phonon context. Early developments
are nicely summarized in the HAC approach.  They present a very simple perturbation theory
derivation that also emphasizes a close connection with self-energy originating from the
first--order coupling (due to translational invariance).

The present approach extends its definition to arbitrary orders and, also, highlights its
physical origin. The Schwinger's variational derivative technique has the merit of showing
that the mean--field potential is due to the dressing of the $\eta$ potential induced by
the $n$th--order fictitious interaction $\xi^{n}$. Physically this corresponds to the
dressing of the electronic potential induced by strongly anharmonic effects.

This also clarifies why the DW potential is not present in any previous
treatment~\cite{hedin_effects_1970,van_leeuwen_first-principles_2004} of the
electron--phonon interaction performed using the Schwinger's variational derivative
technique. The reason is that in these works the e--b interaction is treated at the first
order only.

In conventional theories involving linear electron--boson interactions the $\Langle
\ww{Q}^{n}_{\bmu}\(z_1\)\Rangle$ averages are, in general, connected to the boson mean
displacement ($n=1$) and the population ($n=2$). As a consequence, it is zero for any odd
value of $n$.  The presence of higher--order e--b interactions deeply modifies this simple
scenario. $\Langle \ww{Q}_{\bmu}^n\(z_1\)\Rangle$ is a $n$th--order bosonic tadpole whose
dynamics includes nontrivial contributions, like the one showed in
Fig.\,\ref{fig:DW}(b). These tadpoles are, in general, nonzero.
%
%
%
%
%
\subsection{The mass operator}~\label{sec_mass_operator}
\begin{widetext}
The mass operator requires additional manipulations. We integrate by parts
\begin{figure}[t!]
{\centering\includegraphics[]{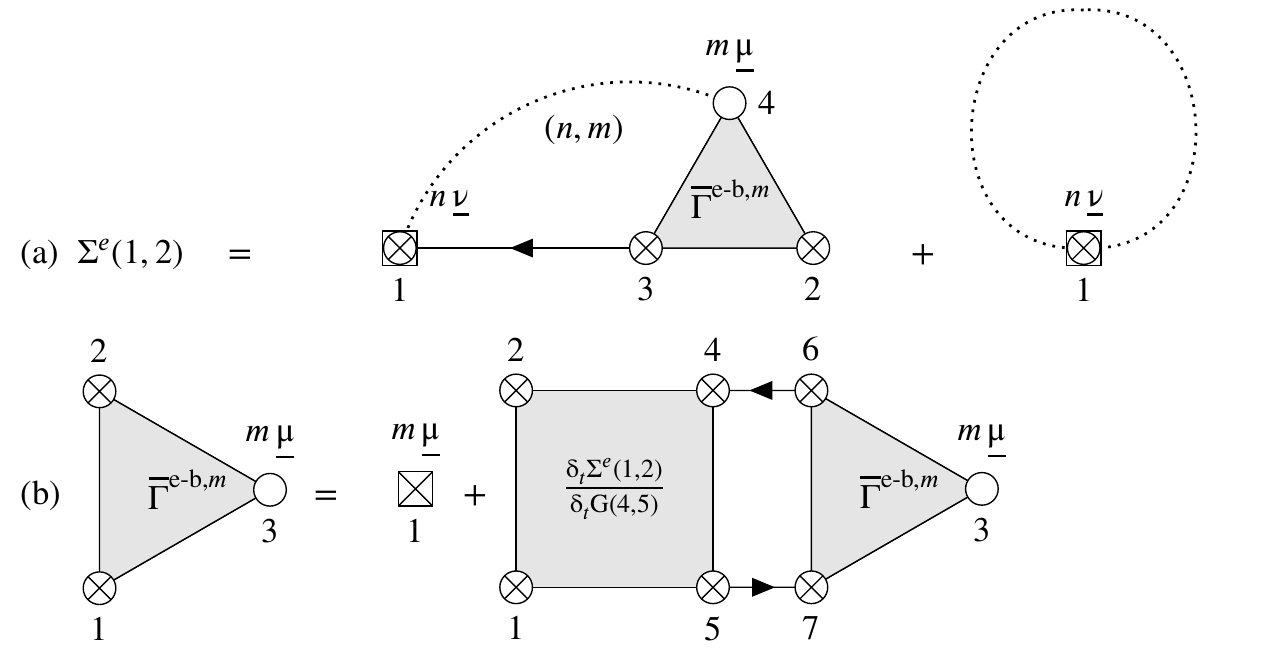}}
  \caption{Diagrammatic form of the self--energy operator (a) and of the vertex function
    (b) for arbitrary orders of the electron--boson interaction and arbitrary number of
    bosons involved in the scattering.  In order to close this set of equations,
    expressions for the bosonic propagator $D^{m,n}$
    (Secs.~\ref{sec_boson_sigma},\ref{sec:responses}) and the vertex function
    $\overline{\Gamma}^{\text{e-b},n}$ (Sec.~\ref{sec_boson_vertexes}) are additionally
    needed. The lowest order approximation for the electron self-energy is described in
    Sec.~\ref{sec_lowest}.}
\label{fig:2}
\end{figure}
\begin{align}
  M\(1,2\)=i\sum_{n,\,\bnu} V_{\bnu}^{n}\(\xx_1\)\int\!\!d3
  \[\frac{\delta}{\delta\xi_{\bnu}^{n}\(z_1\)}G\(1,3\)\]G^{-1}\(3,2\)
  =-i\sum_{n,\,\bnu} V_{\bnu}^{n}\(\xx_1\)\int\!\!d3\,G\(1,3\)
  \frac{\delta}{\delta\xi_{\bnu}^{n}\(z_1\)}G^{-1}\(3,2\).
\end{align}
This equation is exact. Now the problem is how to evaluate this variational derivative. By
noticing that
\begin{align}
\[i\frac{\partial}{\partial z_1}-h_\text{e}\(1\)-\eta\(1\)-\Phi\(1\)\]\delta\(1,2\)=G^{-1}\(1,2\)+M\(1,2\),
\end{align}
we have that
\begin{multline}
  -\frac{\delta}{\delta\xi_{\bnu}^{n}\(z_1\)}G^{-1}\(3,2\)
  =\frac{\delta \Phi\(2\)}{\delta\xi_{\bnu}^{n}\(z_1\)}\delta\(2,3\)
  +\frac{\delta M\(3,2\)}{\delta\xi_{\bnu}^{n}\(z_1\)}
  =\frac{\delta \Phi\(2\)}{\delta\xi_{\bnu}^{n}\(z_1\)}\delta\(2,3\)
  -\int \!\!d4567\,\frac{\delta_t M\(3,2\)}{\delta_t G\(4,5\)}\\
  \times G\(4,6\)\frac{\delta G^{-1}\(6,7\)}{\delta\xi_{\bnu}^{n}\(z_1\)} G\(7,5\).
  \label{eq:def:vertex}
\end{multline} 
\end{widetext}
In \eq{eq:def:vertex} we have introduced the $\delta_t$ symbol to make clear that we are
using a {\em total} derivative.  In this way the derivation of the electronic self--energy
and vertex function closely follows the well--established procedure introduced in the case
of the linear e--b coupling~\cite{van_leeuwen_first-principles_2004}. In the next section
we will further discuss this subtle but important aspect.

We can now define a vertex function that extends to the e--b case the known electronic
vertex function. In order to do so we start by expanding the first term appearing on the
r.h.s. of \eq{eq:def:vertex} using \eq{eq:D_as_dQ_dxi}:
\begin{multline}
  \frac{\delta \Phi\(2\)}{\delta\xi_{\bnu}^{n}\(z_1\)}=\sum_{m,\,\bmu}V_{\bmu}^{m}\(\xx_2\)\label{eq:dvdxi}
  \frac{\delta \Langle Q_{\bmu}^m\(z_2\)\Rangle}{\delta\xi_{\bnu}^{n}\(z_1\)}\\
  =\sum_{m,\,\bmu} V_{\bmu}^{m}\(\xx_2\) D^{m,n}_{\bmu,\bnu}\(z_2,z_1\).
\end{multline}
It is natural to define the electron--boson vertex function, $\Gamma^{\text{e-b},m}_{\bmu}\(1,2;z\)$~\footnote{There is discrepancy in
the literature about the sign in its definition, for instance
Strinati~\cite{strinati_application_1988} defines it with a plus sign, Eq.\,(4.4). Here
we follow a convention in the book of Stefanucci and van
Leeuwen~\cite{stefanucci_nonequilibrium_2013}, Eq.\,(12.34).} as
\begin{multline}
  -\frac{\delta}{\delta\xi_{\bnu}^{n}\(z_1\)}G^{-1}\(3,2\)\equiv\Gamma^{\text{e-b},n}_{\bnu}\(3,2;z_1\)\\
  =-\sum_{m,\,\bmu}\int\!dz_4\,\frac{\delta G^{-1}\(3,2\)}{\delta \Langle Q_{\bmu}^m\(z_4\)\Rangle}
  \frac{\delta \Langle Q_{\bmu}^m\(z_4\)\Rangle}{\delta\xi_{\bnu}^{n}\(z_1\)}\\
=\sum_{m,\,\bmu}\int\!dz_4\,\overline{\Gamma}^{\text{e-b},m}_{\bmu}\(3,2;z_4\) D^{m,n}_{\bmu,\bnu}\(z_4,z_1\).
\label{eq:vertex_def}
\end{multline}
Here, we have also introduced an alternative form of the e--b vertex function:
\begin{align}
  \overline{\Gamma}^{\text{e-b},n}_{\bmu}\(1,2;z_3\)\equiv-\frac{\delta G^{-1}\(1,2\)}{\delta \Langle Q_{\bmu}^n\(z_3\)\Rangle}.
\label{eq:overline_vertex_def}
\end{align}
From \eq{eq:dvdxi} and \eq{eq:overline_vertex_def} it follows that
$\overline{\Gamma}^{\text{e-b}}$ satisfies the following integro-differential equation:
\begin{multline}
  \overline{\Gamma}^{\text{e-b},n}_{\bnu}\(1,2;z_3\)
  =\left.\overline{\Gamma}^{\text{e-b},n}_{\bnu}\(1,2;z_3\)\right|_0+\\
  +\int\!\!d4567\,\frac{\delta_t M\(1,2\)}{\delta_t G\(4,5\)}
  G\(4,6\)\overline{\Gamma}^{\text{e-b},n}_{\bnu}\(6,7;z_3\)G(7,5),\label{eq:vertex}
\end{multline}
with
\begin{multline}
\left.\overline{\Gamma}^{\text{e-b},n}_{\bnu}\(1,2;z_3\)\right|_0
=\frac{\gd \Phi\(1\)}{\gd  \Langle Q_{\bnu}^n\(z_3\)\Rangle}\gd\(1-2\)\\
=\delta\(1-2\)\gd\(z_1-z_3\)V_{\bnu}^{n}\(\xx_1\).
\label{eq:zero_vertex}
\end{multline}
In \eq{eq:vertex} appears $M$ (defined in \eq{eq:sigma_e}) instead of $\Sigma^\text{e}$ as $\eta$ does not depend on
$\xi$ and the lowest order derivative comes through $\Phi$. This, in practice, means that
in the independent particle approximation\,(IPA), ($\Sigma^{\text{e}}=0$), the mixed e--b
vertex is zero, as it should be.

By analogy with electronic case, it can be regarded as the Bethe-Salpeter equation for the
vertex function. It was discussed in the linear electron-phonon coupling by R.~van
Leeuwen~\cite{van_leeuwen_first-principles_2004}. \eq{eq:vertex} also defines the
electron--electron kernel
\begin{align}
K^\text{e-e}\(1,5;2,4\)\equiv\frac{\delta_t \Sigma^\text{e}\(1,2\)}{\delta_t G\(4,5\)},
\label{eq:K_f}
\end{align}
that will also appear in Sec.~\ref{sec:electronic_response} in the case of the equation
of motion for the electronic response function.
Note that in this section we have already introduced a specific notation for the vertex
and for the kernel. Indeed, in both cases we have that the vertex/kernel is defined as the
functional derivative of electronic/bosonic observable (the inverse GF for the vertex and
the self--energy for the kernel) with respect to an electronic/bosonic potential\,(for the
vertex) or GF\,(for the kernel). In the present case $K^\text{e-e}$ is purely electronic,
while in $\Gamma^{\text{e-b}}$ the field, $\xi$, is bosonic.  In the following sections we
will introduce other vertexes and associated kernels and demonstrate that they
are connected via matrix generalization of the Bethe-Salpeter equation.

The full mass operator can be finally written as
\begin{multline}
  M\(1,2\) =i\sum_{n,\,\bnu}\sum_{m,\, \bmu}\int\!\!d3\!\int\!dz_4\,V_{\bnu}^{n}\(\xx_1\) G(1,3)\\
  \times \overline{\Gamma}_{\bmu}^{\text{e-b},m}\(3,2;z_4\) D^{m,n}_{\bmu,\bnu}\(z_4,z_1\).
  \label{eq:Sigma:Gen}
\end{multline}
By comparing the expression for the electron self-energy with the expression in a pure
electronic case one observes that $\sum_{m,\,\bmu}\sum_{n,\,\bnu} V_{\bmu}^{m}(3)
D^{mn}_{\bmu\bnu}\(z_3,z_1\)V_{\bnu}^{n}(1)$ plays the role of the screened Coulomb
interaction.

Eq.~\eqref{eq:Sigma:Gen} is not the most convenient representation of the electron
self-energy because there is no simple way of computing the kernel
$K^\text{e-e}\(1,5;2,4\)$ even though the diagrammatic form of $\Sigma^\text{e}$ is
known. As can be seen from the exact formula, \eq{eq:Sigma:Gen}, and its diagrammatic
representation in Fig.~\ref{fig:2}(a), the self-energy contains the bosonic propagator
$D$, and, therefore, the variation $\frac{\delta \Sigma^\text{e}\(1,2\)}{\delta D\(4,5\)}$
is implicitly included in the $\frac{\delta_t \Sigma^\text{e}\(1,2\)}{\delta_t
  G\(4,5\)}$. This is the main difference from the pure electronic case, where the
screened interaction \emph{explicitly} depends on the electron Green's function. Thus,
although Eq.~\eqref{eq:Sigma:Gen} is exact, it is not practical.  A better approach is to
consider from the beginning the electronic self--energy to be a functional of both
propagators, i.e. $\Sigma^\text{e}=\Sigma^\text{e}\[G,D\]$, which requires the
introduction of other vertex functions. This procedure will be implemented below in
combination with the bosonic self--energy.
%
%
%
%
\section{Single-boson dynamics}~\label{sec_single_boson}
Starting from the equation of motion~\eqref{eom:Q} for $\ww{Q}_\mu$ we derive the equation
of motion for the bosonic propagator $D_{\mu,\nu}$ in a similar way to the electronic
case:
\begin{widetext}
\begin{multline}
 -\frac{1}{\Omega_\mu}\[\frac{\partial^2}{\partial z_1^2}+\Omega_\mu^2\]D_{\mu,\nu}\(z_1,z_2\)=
  \delta_{\mu\nu}\delta\(z_1-z_2\)\\
-i\sum_{\bgz,n} n 
\Bigg[\int\!\!d\xx_1\, V^{n}_{\bgz\oplus\mu}\(\xx_1\)
\underbrace{\La\mathcal{T}  \Delta\[\hat \gr\(1\)\, \hat Q^{n-1}_{\bgz}\(z_1\)\]\hat Q_\nu\(z_2\)\Ra}_{J^{\(n\)}_V}+
\xi^n_{\bgz\oplus\mu}\(z_1\)\underbrace{\La\mathcal{T} \Delta\hat Q^{n-1}_{\bgz}\(z_1\)\hat Q_\nu\(z_2\)\Ra}_{J^{\(n\)}_\xi}\Bigg],
\label{eom:D_1}
\end{multline}
\end{widetext}
The last term is driven  by the auxiliary fields $\xi^{n}_{\bgz}$. According to the rules
specified above, the limit of zero $\xi^n_{\bgz}$ is to be taken at the end of
derivations. 

In \eq{eom:D_1} we have schematically represented with $J^{\(n\)}_V$ and $J^{\(n\)}_\xi$,
respectively, the term induced by the scattering potential and by the auxliary field. The
goal of this section is to rewrite {\em exactly}
\begin{multline}
  -i\sum_{\bgz,n} n \[\int\!\!d\xx_1\, V^{n}_{\bgz\oplus\mu}\(\xx_1\)J^{\(n\)}_V
  +\xi^n_{\bgz\oplus\mu}\(z_1\)J^{\(n\)}_\xi\]\\
  =\sum_{\ga}\int \!dz_3 \[\Pi_{\mu,\ga}\(z_1,z_3\)\capo
  + \(U_{\mu,\ga}\(z_1\)+\Xi_{\mu,\ga}\(z_1\)\)\gd\(z_1-z_3\)\]\\
  \times D_{\ga,\nu}\(z_3,z_2\).
\label{eom:plan}
\end{multline}
In \eq{eom:plan} we have introduced the generalized bosonic mass operator, $\Pi$ and
the mean--field potentials, $U$ and $\Xi$. $\Xi$ is driven by the fictitious external field
and vanishes when $\xi\rightarrow 0$.  $\Pi$, $U$ and $\Xi$ sum in the total bosonic
self--energy $\Sigma^\text{b}$ that, consistently with \eq{eq:sigma_e}, is defined as
\begin{align}
  \Sigma^\text{b}_{\mu,\nu}\(z_1,z_2\)=\Pi_{\mu,\nu}\(z_1,z_2\) + U_{\mu,\nu}\(z_1\)\gd\(z_1-z_2\).
  \label{eq:sigma_b}
\end{align}
In order to find the explicit expression for $\Pi$, $U$ and $\Xi$, we start by observing
that \eq{eom:D_1} includes linear\,($n=1$) and higher-order ($n>1$) terms. In the $n=1$
case, $\langle \hat Q\rangle=0$, and we can use the chain rule to write
\begin{multline}
  J^{\(1\)}_V=\La\mathcal{T} \hat \gr\(1\)\hat Q_\nu\(z_2\) \Ra
  =i\frac{\gd \La \hat \gr\(1\) \Ra}{\gd \xi^{1}_{\nu}\(z_2\)}\\=
-\int\!d34\, G\(1,3\) \frac{\delta G^{-1}(3,4)}{\delta\xi^{1}_{\nu}\(z_2\)} G\(4,1\).
\label{eom:M_1_1}
\end{multline}
We can now use the definition of the electronic vertex, \eq{eq:vertex_def}, and rewrite
$J^{\(1\)}_V$ in terms of the mass operator $\Pi^1$:
\begin{multline}
  \Pi^{1}_{\mu,\ga}\(z_1,z_2\) = \int\!d34\int\!d\xx_1\, V^{1}_{\mu}\(\xx_1\)\\
  G\(1,3\) G\(4,1\) \overline{\Gamma}^{\text{e-b},1}_{\ga}\(3,4;z_2\),
  \label{eom:M_1_4}
\end{multline}
that is diagrammatically represented in Fig.\,(\ref{tbl:D_SEs}a).  This contribution to the
bosonic mass operator does not require further manipulations and is explicit function of
the single--boson correlator $D$.  $\Pi^{\(1\)}$ represents the
generalization to the case of non--linear e-b coupling of the first--order e-b
mass operator well known and widely used in the
literature~\cite{Mahan1990,ALEXANDERL.FETTER1971} to calculate, for example, phonon
linewidths~\cite{PhysRevB.91.054304}.

We now move to the $n>1$ case. We observe that, thanks to \eq{eq:Q0},
\begin{align}
\La\mathcal{T} \Delta\hat Q^{n-1}_{\bgz}\(z_1\)\hat Q_\nu\(z_2\)\Ra=i D^{n-1,1}_{\bgz,\nu}\(z_1,z_2\),
\end{align}
and, by using \eq{eq:D_mm1_1_as_dD} with $m=n$ and $k=n-2$  we can express $J^{\(n>1\)}_\xi$ as
\begin{multline}
  J^{\(n>1\)}_\xi=\sum_{\bgz} n\,\xi^n_{\bgz\oplus\mu}\(z_1\) D^{n-1,1}_{\bgz,\nu}\(z_1,z_2\)\\
  =\sum_{\bku,\alpha} n\,\xi^n_{\mu\oplus\bku\oplus\alpha}\(z_1\)\(\Langle\ww{Q}^{n-2}_{\bku}\(z_1\)\Rangle
  +i\frac{\delta}{\delta \xi^{n-2}_{\bku}\(z_1\)}\)\\ \times D_{\alpha,\nu}\(z_1,z_2\).\label{eq:def:Phi}
\end{multline}
with $\bgz=\bku\oplus\alpha$.

The $J^{\(n>1\)}_V$ correlator can be evaluated by using \eq{id:diff_prime}:
\begin{multline}
  J^{\(n>1\)}_{V}=
  \(\Langle\ww{Q}^{n-2}_{\bku}\(z_1\)\Rangle +i\frac{\delta}{\delta \xi^{n-2}_{\bku}\(z_1\)}\)\\\times
  \(\Langle\hat \rho(1)\Rangle+i\frac{\delta}{\delta \eta\(1\)}\) D_{\alpha,\nu}\(z_1,z_2\).
\label{eq:J_final}
\end{multline}
In \eq{eq:J_final} the $\gd \eta$ derivative is made acting before the $\gd \xi$ one. In
this way the limit of zero external field can be safely taken and the last term of
\eq{eq:D_mm1_1_as_dD} vanishes. It is, indeed, important to remind that
$\Langle\ww{Q}_{\ga}\(z_1\)\Rangle=0$ only when $\xi=0$.

If we now collect \eq{eq:def:Phi} and \eq{eq:J_final} and plug them in \eq{eom:D_1} we can
recast the EOM for $D$ in the form
\begin{widetext}
\begin{multline}
 -\frac{1}{\Omega_\mu}\Big[\frac{\partial^2}{\partial z_1^2}+\Omega_\mu^2\Big]D_{\mu,\nu}\(z_1,z_2\)=
  \delta_{\mu\nu}\delta\(z_1-z_2\)
  +\sum_{n>1,\,\bku,\alpha}  n\,
\underbrace{
 \(
\langle\hat\gamma^n_{\mu\oplus\bku\oplus\alpha}\(z_1\)\rangle+\xi^n_{\mu\oplus\bku\oplus\alpha}\(z_1\)\)
  \La\ww{Q}_{\bku}^{n-2}\(z_1\)\Ra D_{\alpha,\nu}\(z_1,z_2\)}_{U+\Xi}\\  
  +\sum_{\alpha}\int\! dz_3\, \Pi^{\(1\)}_{\mu,\ga}\(z_1,z_3\)D_{\ga,\nu}\(z_3,z_2\)\\
  +\sum_{n>1,\,\bku,\alpha}  n\,\int\!\!d\xx_1\, V^{n}_{\mu\oplus\bku\oplus\alpha}\(\xx_1\) 
   \Bigg[ 
     \underbrace{
       i\frac{\gd \[\gr\(1\)D_{\alpha,\nu}\(z_1,z_2\)\]}{\gd \xi^{n-2}_{\bku}\(z_1\)}}_{\Pi^{\(2\)}}
     +\underbrace{i\La\ww{Q}_{\bku}^{n-2}\(z_1\)\Ra\frac{\gd D_{\alpha,\nu}\(z_1,z_2\)}{\gd\eta\(1\)}}_{\Pi^{\(3\)}}
     -\underbrace{\frac{\gd^2D_{\alpha,\nu}\(z_1,z_2\)}{\gd\eta\(1\)\gd\xi^{n-2}_{\bku}\(z_1\)}}_{\Pi^{\(4\)}}\Bigg].
 \label{eom:D_2}
\end{multline}
\end{widetext}
Eq.~\eqref{eom:D_2} represents a key result of this work. We have already schematically
identified the different terms that compose the EOM for $D$. The $J^{\(n\)}_\xi$ term
reduces, when $\xi^n\rightarrow 0$ only to the $U$ potential, while the $J^{\(n\)}_V$ term
reduces to the sum of three mass operators. In the following we study them in detail in
order to recast \eq{eom:D_2} in the form of a Dyson equation for $D$.
%
%
%
%
%
\subsection{Mean-field potentials}~\label{sec_boson_mean_field}
\begin{figure}[t!]
 \center
 \includegraphics[]{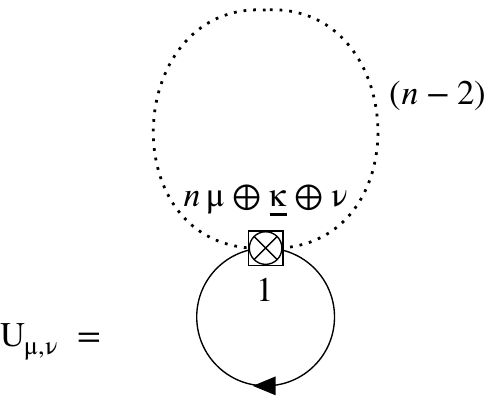}
 \caption{The $n$th--order bosonic mean--field potential is one of the constituents of the
   total bosonic self--energy.
 \label{fig:SE_D_2}}  
\end{figure}

The first contribution to the EOM for $D$ is through the mean--field potentials, $U$ and
$\Xi$. These potentials are due to the first term on the r.h.s. of \eq{eq:def:Phi} and to
the $\Langle\ww{Q}^{n-2}_{\bku}\(z_1\)\Rangle D_{\alpha,\nu}\(z_1,z_2\)$ term in
\eq{eom:D_2}. The sum of these two terms can be rewriten as the action of two local
potentials on the bosonic propagator:
\begin{equation}
 \sum_{\alpha} \[U_{\mu,\ga}\(z_1\)+\Xi_{\mu,\ga}\(z_1\)\]  D_{\alpha,\nu}\(z_1,z_2\),
  \label{eq:U_1}
\end{equation} 
with
\begin{subequations}
\begin{gather}
  U_{\mu,\alpha}\(z_1\)=\sum_{n \geqslant 2,\,\bku} n
  \langle\hat\gamma^n_{\mu\oplus\bku\oplus\alpha}\(z_1\)\rangle\Langle\ww{Q}^{n-2}_{\bku}\(z_1\)\Rangle,\label{eq:U_2}\\
  \Xi_{\mu,\alpha}\(z_1\)=\sum_{n \geqslant 2,\,\bku} n \xi^n_{\mu\oplus\bku\oplus\alpha}\(z_1\)
  \Langle\ww{Q}^{n-2}_{\bku}\(z_1\)\Rangle.
\end{gather} 
\label{eq:U_2p}
\end{subequations}
We remind the reader that
$\xi^n_{\mu\oplus\bku\oplus\alpha}$ and $\hat\gc^{n}_{\mu\oplus\bku\oplus\alpha}$ are
symmetric tensors of rank $n$, and $\bku$ is an $n-2$ dimensional
vector. Eq.\,\eqref{eq:U_2} is represented diagrammatically in \fig{fig:SE_D_2} in the
limit of vanishing auxiliary fields.

%
\subsection{The pure bosonic vertex function $\Gamma^\text{b-b}$}~\label{sec_bb_vertex}
A key ingredient of \eq{eom:D_2} is the first order derivative $\frac{\delta
  D\(z_1,z_2\)}{\delta \xi^{n}\(z_1\)}$. This term shows some remarkable properties that
we study here in detail.  We start from the term
\begin{multline}
  \frac{\delta D_{\alpha,\nu}\(z_1,z_2\)}{\delta \xi^{n-2}_{\bku}\(z_1\)}= \sum_{\beta,\gamma}
  \int\! dz_3 dz_4\, D_{\alpha,\beta}\(z_1,z_3\)\\
\times\Gamma^{\text{b-b},n-2}_{\beta,\gamma;\bku}\(z_3,z_4;z_1\) D_{\gamma,\nu}\(z_4,z_2\).
\label{eq:b_resp.2:duplicate}
\end{multline}
Eq.~\eqref{eq:b_resp.2:duplicate} introduces a further vertex with an entire bosonic
character:
\begin{align}
  \Gamma^{\text{b-b},n}_{\beta,\gamma;\bku}\(z_1,z_2;z_3\)\equiv
  - \frac{\delta D^{-1}_{\beta,\gamma}\(z_1,z_2\)}{\delta \xi^{n}_{\bku}\(z_3\)}.
\label{eq:b_resp.3:duplicate}
\end{align}
The lowest--order contribution to this vertex function is from the variational derivative
of the driving field entering the mean--field potential, \eq{eq:U_2}:
\begin{multline}
  \left.\Gamma^{\text{b-b},n}_{\beta,\gamma;\bgl}\(z_1,z_2;z_3\)\right|_0 =  \\ 
  \sum_{m \geqslant 2,\,\bku} m \frac{\delta \[ \xi^m_{\mu\oplus\bku\oplus\alpha}\(z_1\)
    \Langle\ww{Q}^{m-2}_{\bku}\(z_1\)\Rangle\] }{\delta \xi^{n}_{\bgl}\(z_3\)} \gd\(z_1-z_2\).
\label{eq:v_bb.1}
\end{multline}
In the limit $\xi\rightarrow 0$ only the derivative of
$\xi^m_{\mu\oplus\bku\oplus\alpha}\(z_1\)$ gives a nonzero contribution. As written
previously the $\xi$ function is totally symmetric. In practice this means that, if we
call $\bgI$ the $n$--dimensional vector containing a generic permutation of the $\bgl$
indexes, we have that $\xi^n_{\bgl}=\xi^{n}_{\gl_{I_1},\dots,\gl_{I_n}}$. It follows that
\begin{multline}
\frac{\delta \xi^m_{\mu\oplus\bku\oplus\alpha}\(z_1\)}{\delta \xi^{n}_{\bgl}\(z_3\)}=\gd\(z_1-z_3\) \gd_{nm}\\
 \times\frac{n}{n!} \sum_{I=1}^{n!}  \gd_{\kappa_1,\gl_{I_2}}\cdots\gd_{\kappa_1,\gl_{I_{n-1}}} 
 \gd_{\mu,\gl_{I_1}}\gd_{\alpha,\gl_{I_n}}.
\label{eq:v_bb.2}
\end{multline}
Eq.~\eqref{eq:v_bb.2} gives, in practice, only $n\(n-1\)$ terms as all $\binom{n}{n-2}$
permutations of $\gl_{I_2}\dots\gl_{I_{n-1}}$ inside the
$\Langle\ww{Q}^{m-2}_{\gl_{I_2}\dots\gl_{I_{n-1}}}\(z_1\)\Rangle$ gives the same
contribution. The final form of $\left.\Gamma^{\text{b-b},n}\right|_0$ is, therefore:
\begin{multline}
  \left.\Gamma^{\text{b-b},n}_{\beta,\gamma;\bgl}\(z_1,z_2;z_3\)\right|_0\equiv
 \left.\Gamma^{\text{b-b},n}_{\beta,\gamma;\bgl}\(z_1\)\right|_0\gd\(z_1-z_2\)\gd\(z_1-z_3\)\\=
  \frac{1}{\(n-1\)!} \sum_{I=1}^{n!} \Langle\ww{Q}^{n-2}_{\gl_{I_2}\dots\gl_{I_{n-1}}}\(z_1\)\Rangle\\\times
\gd_{\mu,\gl_{I_1}}\gd_{\alpha,\gl_{I_n}}\gd\(z_1-z_2\)\gd\(z_1-z_3\).
\label{eq:v_bb.3}
\end{multline}
Note the contracted single--time form of $\left.\Gamma^{\text{b-b},n}\right|_0$ introduced
in \eq{eq:v_bb.3}. It will be used in the zeroth--order approximations for
$\Sigma^{\text{b}}$, cf. Eq.~\eqref{eq:sigma_b}.
%
%
%
%
%
\subsection{Nonlinear self--energies }~\label{sec_boson_sigma}
The first term we analyze is $\Pi^{\(2\)}$. With the help of \eq{eq:b_resp.2:duplicate},
it follows that
\begin{multline}
  \Pi^{\(2a\)}_{\mu,\nu}\(z_1,z_2\)
  =i\sum_{n \geqslant 2,\,\bku,\ga} n \langle\hat\gamma^n_{\mu\oplus\bku\oplus\alpha}\(z_1\)\rangle\\
  \times\sum_{\gb} \int\! dz_3\, D_{\alpha,\beta}\(z_1,z_3\)\Gamma^{\text{b-b},n-2}_{\beta,\nu;\bku}\(z_3,z_2;z_1\).
  \label{eq:P:1}
\end{multline}
By expressing the electron density in terms of the equal times Green's function as
$\gr\(1\)=-iG\(1,1^+\)$, we compute the variation $\frac{\delta
  \Langle\hat\rho\(1\)\Rangle}{\delta \xi^{n-2}_{\bku}\(z_1\)}$. It yields
\begin{widetext}
\begin{align}
  \Pi^{\(2b\)}_{\mu,\nu}\(z_1\)=\sum_{n\geqslant 2,\,\bku,\bgl,l} n \, \int\! d\xx_1\,V^n_{\mu\oplus\bku\oplus\nu}\(\xx_1\)
  \int\!d34\,G(1,3) \int\!dz_5\, \oo{\Gamma}^{\text{e-b},l}_{\blu}\(3,4;z_5\)D^{l,n-2}_{\blu,\bku}\(z_5,z_1\)G\(4,1^+\).
  \label{eq:U:1}
\end{align}  
This mass operator is \emph{local} and can be seen as a correlated correction to $U$.
There is no analogous contribution to the mean--field potential in pure electronic
systems, and to the best of our knowledge, it was not discussed in the context of e-b
interactions.

Next we consider the $\Langle\ww{Q}^{n-2}_{\bku}\Rangle\frac{\gd D}{\gd\eta}$ variation 
\begin{align}
  \Pi^{\(3\)}_{\mu,\nu}\(z_1,z_2\)
  =i\sum_{n \geqslant 2,\,\bku,\alpha} n\, \int\! d\xx_1\,V^n_{\mu\oplus\bku\oplus\alpha}\(\xx_1\)
  \Langle\ww{Q}^{n-2}_{\bku}\(z_1\)\Rangle
  \sum_{\beta} \int\! dz_3\, D_{\alpha,\beta}\(z_1,z_3\)\Gamma^{\text{b-e}}_{\beta,\nu}\(z_3,z_2;1\),
  \label{eq:P:2}
\end{align}
where we used the chain rule of differentiation and introduced a new vertex function with
two bosonic and one fermionic coordinates:
\begin{align}
   \Gamma^{\text{b-e}}_{\beta,\gamma}\(z_1,z_2;3\)\equiv
  - \frac{\delta D^{-1}_{\beta,\gamma}\(z_1,z_2\)}{\delta \eta\(3\)}
  = - \int\!d4\,\frac{\delta D^{-1}_{\beta,\gamma}\(z_1,z_2\)}{\delta\Langle\hat\rho\(4\)\Rangle}
  \frac{\delta\Langle\hat\rho\(4\)\Rangle}{\delta \eta\(3\)}
  \equiv\int\!d4\,\overline{\Gamma}^{\text{b-e}}_{\beta,\gamma}\(z_1,z_2;4\)\chi\(4,3\).
 \label{eq:e_resp.0a}
\end{align}
Notice, that similarly to the other mixed vertex,~\eq{eq:vertex_def}, we pulled out the
common part of the functional derivative from the definition. The common part is given by
the electron density response function
\begin{align}
  \chi(1,2)=\frac{\delta \langle \hat \gr\(1\)\rangle}{\delta \eta\(2\)}
  =-i\frac{\delta G\(1,1^+\)}{\delta \eta\(2\)}.
 \label{eq:e_resp.1}
\end{align}
Other terms as well as contributions to the vertex function
$\Gamma^{\text{b-e}}_{\beta,\gamma}\(z_1,z_2;3\)$ from the bosonic self--energy will be
considered in the next section.

Our next contribution results from the application of double differential operators
$\frac{\delta^2}{\delta \xi^{n-2}_{\bku}\(z_1\)\delta\eta\(1\)}$ and consists of three
terms
\begin{subequations}
\label{eq:P:3}
\begin{multline}
  \Pi^{\(4a\)}_{\mu,\nu}\(z_1,z_2\) =-\sum_{n \geqslant 2,\,\bku, \alpha} n \, \int\! d\xx_1\,V^n_{\mu\oplus\bku\oplus\alpha}\(\xx_1\)
  \sum_{\beta,\phi,\psi} \int\! dz_3 dz_4 dz_5\, D_{\alpha,\phi}\(z_1,z_4\)\\\times
  \Gamma^{\text{b-b},n-2}_{\phi,\psi;\bku}\(z_4,z_5;z_1\) D_{\psi,\beta}\(z_5,z_3\)\Gamma^{\text{b-e}}_{\beta,\nu}\(z_3,z_2;1\),
\end{multline}
\begin{multline}
 \Pi^{\(4b\)}_{\mu,\nu}\(z_1,z_2\)=-\sum_{n \geqslant 2,\,\bku, \alpha} n \, \int\! d\xx_1\,V^n_{\mu\oplus\bku\oplus\alpha}\(\xx_1\) 
 \sum_{\beta,\phi,\psi} \int\! dz_3 dz_4 dz_5\, D_{\alpha,\beta}\(z_1,z_3\)\\\times\Gamma^{\text{b-e}}_{\beta,\phi}\(z_3,z_4;1\)
 D_{\phi,\psi}\(z_4,z_5\)\Gamma^{\text{b-b},n-2}_{\psi,\nu;\bku}\(z_5,z_2;z_1\),
\end{multline}
\begin{align}
 \Pi^{\(4c\)}_{\mu,\nu}\(z_1,z_2\)+\Pi^{\(4d\)}_{\mu,\nu}\(z_1,z_2\)=-\sum_{n \geqslant 2,\,\bku, \alpha} n \, \int\! d\xx_1\,V^n_{\mu\oplus\bku\oplus\alpha}\(\xx_1\) 
 \sum_{\beta} \int\! dz_3\, D_{\alpha,\beta}\(z_1,z_3\)
 \frac{\delta\Gamma^{\text{b-e}}_{\beta,\nu}\(z_3,z_2;1\)}{\delta \xi^{n-2}_{\bku}\(z_1\)}.
 \label{eq:P:3c}
\end{align}
\end{subequations}
Note that \eq{eq:P:3c} produces two terms, $\Pi^{\(4c\)}$ and $\Pi^{\(4d\)}$ as it will be
demonstrated in the next section.
\begin{figure}[t!]
  {\centering\includegraphics[scale=0.7]{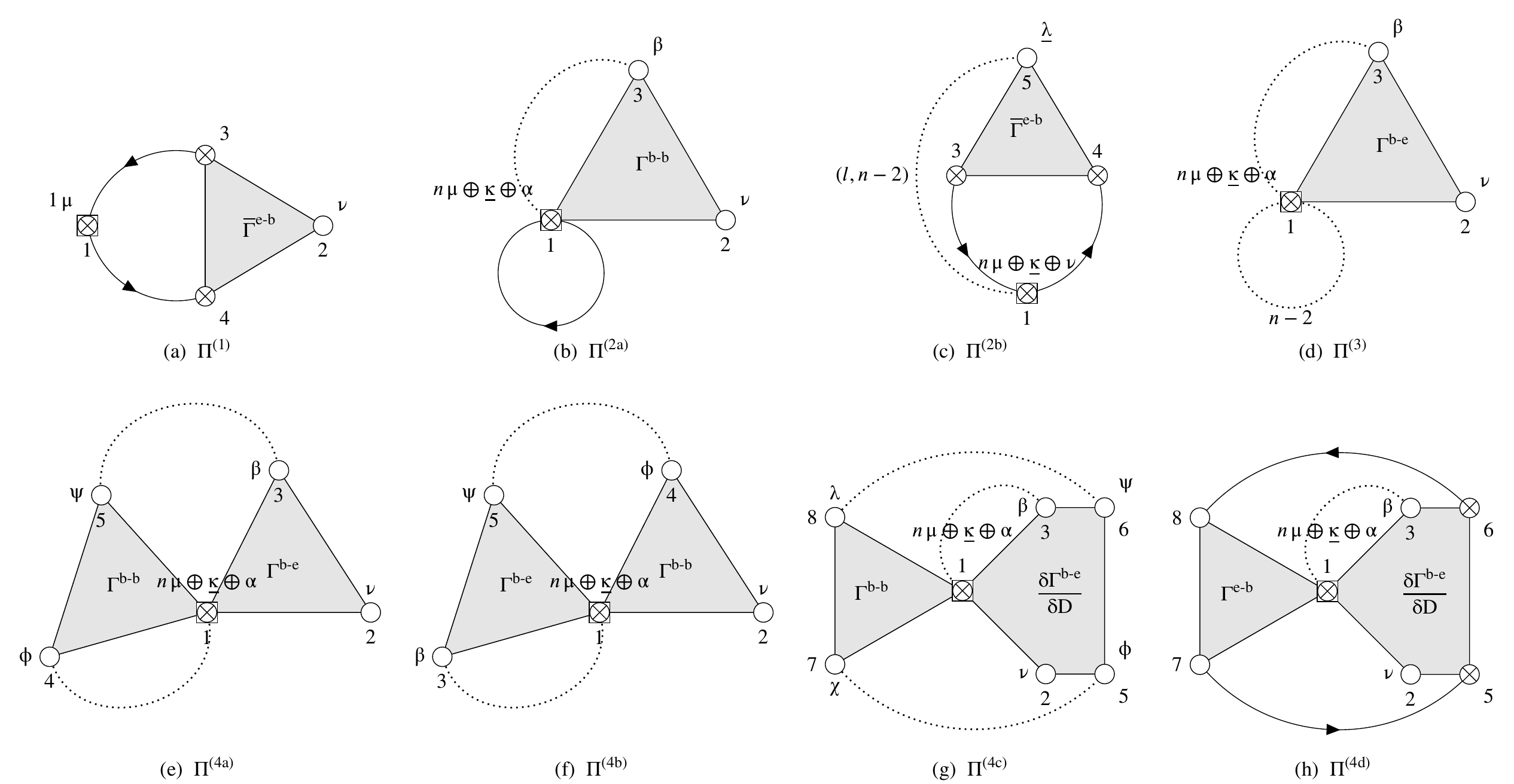}}
  \caption{A total of eight diagrams constituting the exact bosonic mass operator
    $\Pi_{\mu,\nu}(z_1,z_2)$.\label{tbl:D_SEs}}
\end{figure}
%
%
%
%
%
\subsection{Vertex functions}\label{sec_boson_vertexes}
In the preceeding sections we derived the equation of motion of the bosonic propagator
$D_{\mu,\nu}$, \eq{eom:D_2}. Its important ingredients are the mean--field potentials
$U_{\mu,\nu}$ and $\Xi_{\mu,\nu}$, \eq{eq:U_2} and \eq{eq:U_2p} and the bosonic mass
operator $\Pi$ consisting of eight terms $\Pi^{\(1\)}$, $\Pi^{\(2a\)}$, $\Pi^{\(2b\)}$,
$\Pi^{\(3\)}$, $\Pi^{\(4a\)}$, $\Pi^{\(4b\)}$, $\Pi^{\(4c\)}$ and $\Pi^{\(4d\)}$.  They,
in turn, explicitly depend on three vertex functions: $\Gamma^{\text{e-b}}$,
$\Gamma^{\text{b-e}}$, and $\Gamma^{\text{b-b}}$.  $\Gamma^{\text{e-e}}$ appears implictly
through the response function $\chi$, in \eq{eq:e_resp.0a}. The vertex functions contain
one, two or three external bosonic indices.  In order to close the functional equations,
we still need to express these vertex functions in terms of already defined correlators.

In order to do so, let us rewrite the vertex function as components of a \emph{Jacobian
  matrix}:
\begin{align}  
  \vec{\Gamma}\(1,2;3\)\equiv-\[
  \def\arraystretch{2}
  \begin{array}{cc}
    \frac{\delta G^{-1}\(1,2\)}{\delta\eta\(3\)}&
    \frac{\delta G^{-1}\(1,2\)}{\delta\xi_{\bku}^{n}\(z_3\)}\\
    \frac{\delta D^{-1}_{\mu,\nu}\(z_1,z_2\)}{\delta\eta\(3\)}&
    \frac{\delta D^{-1}_{\mu,\nu}\(z_1,z_2\)}{\delta\xi_{\bku}^{n}\(z_3\)}
  \end{array}  
  \]
=\[\begin{array}{cc}
    \Gamma^{\text{e-e}}\(1,2;3\)&
    \Gamma_{\bku}^{\text{e-b},{n}}\(1,2;z_3\)
    \vphantom{\frac{\delta G^{-1}\(1,2\)}{\delta\xi_{\bku}^{n}\(z_3\)}}\\
    \Gamma^{\text{b-e}}_{\mu,\nu}\(z_1,z_2;3\)&
    \Gamma^{\text{b-b},{n}}_{\mu,\nu;\bku}\(z_1,z_2;z_3\)
    \vphantom{\frac{\delta D^{-1}_{\mu,\nu}\(z_1,z_2\)}{\delta\xi_{\bku}^{n}\(z_3\)}}
  \end{array}  
  \],\label{eq:matrix:J}
\end{align}
and 
\begin{align}
  \vec{K}\(1,5;2,4\)\equiv\[
  \def\arraystretch{2}
  \begin{array}{cc}
    \frac{\delta \Sigma^\text{e}\(1,2\)}{\delta G\(4,5\)}&
    \frac{\delta M\(1,2\)}{\delta D_{\phi,\psi}\(z_4,z_5\)}\\
    \frac{\delta \Pi_{\mu,\nu}\(z_1,z_2\)}{\delta G\(4,5\)}&
    \frac{\delta \Sigma^\text{b}_{\mu,\nu}\(z_1,z_2\)}{\delta D_{\phi,\psi}\(z_4,z_5\)}
  \end{array}  
  \]
=\[\begin{array}{cc}
    K^\text{e-e}\(1,5;2,4\)&
    K^\text{e-b}\(1,z_5;2,z_4\)\vphantom{\frac{\delta M\(1,2\)}{\delta D_{\phi,\psi}\(z_4,z_5\)}}\\
    K^\text{b-e}\(z_1,5;z_2,4\)&
    K^\text{b-b}\(z_1,z_5;z_2,z_4\)\vphantom{ \frac{\delta \Pi_{\mu,\nu}\(z_1,z_2\)}{\delta D_{\phi,\psi}\(z_4,z_5\)}}
  \end{array}  
  \].
\label{eq:matrix:K}
\end{align}
Here, $\vec \Gamma$ is built of the vertex functions, and $\vec K$ is the matrix of kernels.

The definitions introduced with \eq{eq:matrix:J} and \eq{eq:matrix:K} make clear that the
electronic and bosonic degrees of freedom are totally symmetric and treated on equal
footing. Indeed the rows and columns of the two matrices can be labelled with the kind of
input/output legs of the vertex/kernel $\[
  \begin{array}{ll}
     \text{e-e}&
     \text{e-b}\\
     \text{b-e}&
     \text{b-b}
  \end{array}  
  \]$.
\begin{figure}[t!]
  {\centering
  \includegraphics[width=\columnwidth]{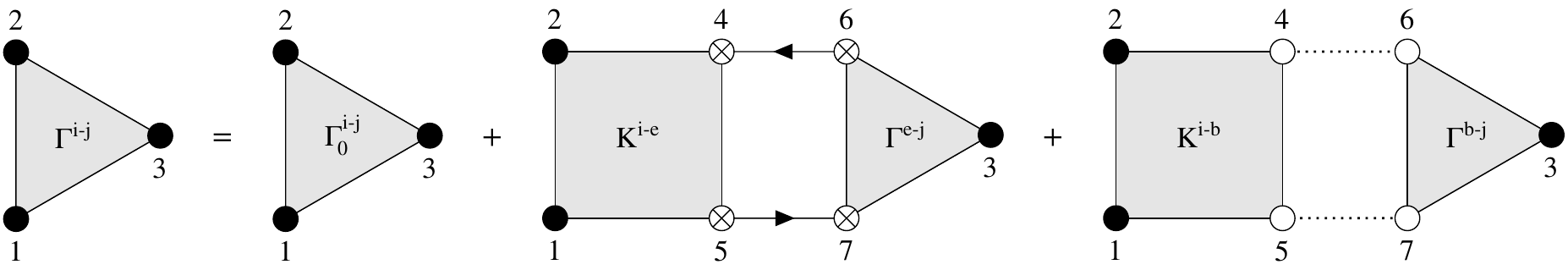}
  }
  \caption{Diagrammatic form of the generalized Bethe--Salpeter equation. Black dots
    denote generic electron or boson indexes, $i,\,j=\(\text{e},\,\text{b}\)$.}
\label{fig:BSE_general}  
\end{figure}
\begin{figure}[t!]
  {\centering
  \includegraphics[]{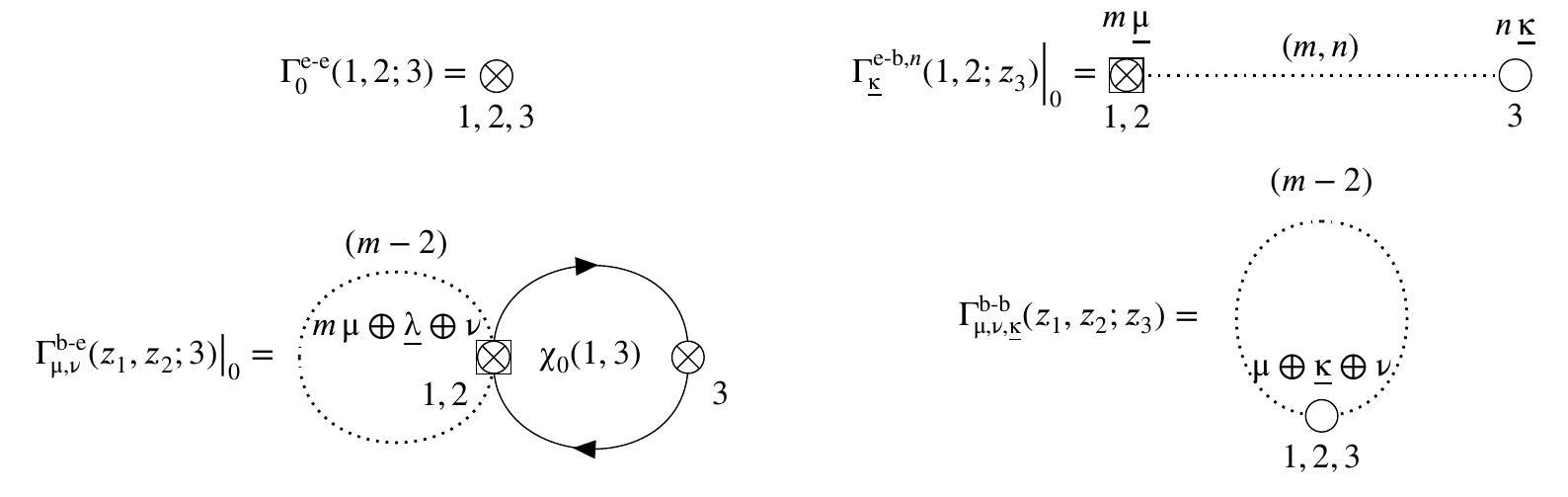}
  }
  \caption{Diagrammatic representation of the lowest-order vertex functions,
    \eq{eq:J0}.}.\label{fig:Gamma_0}
\end{figure}

For a given diagrammatic expression of the electronic and bosonic self-energies, the
corresponding \emph{partial} variations can be easily computed.  Finally, we introduce the
free term given by the derivatives of the mean-field electronic and bosonic potentials:
\begin{align}
  \def\arraystretch{1.5}
  \vec{\Gamma_0}\(1,2;3\)\equiv\[
  \begin{array}{ll}
     \Gamma_0^{\text{e-e}}\(1,2;3\)&
     \Gamma_0^{\text{e-b}}\(1,2;z_3\)\\
     \Gamma_0^{\text{b-e}}\(z_1,z_2;3\)&
     \Gamma_0^{\text{b-b}}\(z_1,z_2;z_3\)
  \end{array}  
  \]\equiv\[
  \begin{array}{ll}
     \frac{\gd \eta\(1\)}{\gd \eta\(3\)} \gd\(1-2\)&
     \frac{\gd \Phi\(1\)}{\gd \xi^{n}_{\bku}\(z_3\)} \gd\(1-2\)\\
     \frac{\gd U_{\mu,\nu}\(z_1\)}{\gd \eta\(3\)} \gd\(z_1-z_2\)&
     \frac{\gd \Xi_{\mu,\nu}\(z_1\)}{\gd \xi^{n}_{\bku}\(z_3\)} \gd\(z_1-z_2\)
  \end{array}  
  \],
\label{eq:matrix:J0}
\end{align}
with
\begin{subequations}
  \label{eq:J0}
\begin{gather}
 \Gamma_0^{\text{e-e}}\(1,2;3\)=\delta(1-2)\delta(1-3),\\
 \Gamma_0^{\text{e-b}}\(1,2;z_3\)=\left. \Gamma_{\bku}^{\text{e-b},{n}}\(1,2;z_3\)\right|_0
 =\sum_{m,\,\bmu} V_{\bmu}^{m}\(\xx_1\) D^{m,{n}}_{\bmu,\bku}\(z_1,z_3\)\delta(1-2),\\
 \Gamma_0^{\text{b-e}}\(z_1,z_2;3\)=\left. \Gamma^{\text{b-e}}_{\mu,\nu}\(z_1,z_2;3\)\right|_0 = 
 \sum_{m,\,\blu} m \Langle \hat Q_{\blu}^{m-2}\(z_1\)\Rangle
 \int \!d\xx_1 \,V_{\mu\oplus\blu\oplus\nu}^{m}\(\xx_1\)\chi\(1,3\)\delta(z_1-z_2),\label{eq:G0.be}\\
 \Gamma_0^{\text{b-b}}\(z_1,z_2;z_3\)=  \left. \Gamma_{\mu,\nu;\bku}^{\text{b-b},{n}}\(z_1,z_2;z_3\)\right|_0= 
 \frac{1}{\(n-1\)!} \sum_{I=1}^{n!} \Langle\ww{Q}^{n-2}_{\kappa_{I_2}\dots\kappa_{I_{n-1}}}\(z_1\)\Rangle
 \gd_{\mu,\kappa_{I_1}}\gd_{\nu,\kappa_{I_n}}\gd\(z_1-z_2\)\gd\(z_1-z_3\).\label{eq:G0.bb}
\end{gather}
\end{subequations}
These four quantities are related by a system of linear equations:
\begin{multline}
\Gamma^{\text{i-j}}\(1,2;3\)= 
\Gamma^{\text{i-j}}_0\(1,2,3\)+ K^{\text{i-e}}\(1,5;2,4\) G\(4,6\) G\(7,5\) \Gamma^{\text{e-j}}\(6,7;3\)\\+
K^{\text{i-b}}\(1,5;z_2,z_4\) D_{\psi,\xi}\(z_4,z_6\) D_{\phi,\eta}\(z_7,z_5\) \Gamma^{\text{b-j}}_{\xi,\phi}\(z_6,z_7;3\),
\label{eq:gen:BSE}
\end{multline}
where the summation and the integration over the repeated arguments is assumed, and the
generic indexes are $i,\,j=\(\text{e, b}\)$. This is the sought \emph{generalized
  Bethe-Salpeter equation}\,(GBSE) for the vertex functions.

Now we are in the position to evaluate Eq.~\eqref{eq:P:3c}, which, in fact, contains the
variation $\frac{\delta \vec{\Gamma}^{\text{b-e}}}{\delta \xi^{n}}$. Since $\vec \Gamma$
is a solution of the complicated equation, its explicit form is not known. Therefore we
use again the chain rule:
\begin{align}
  \frac{\delta \Gamma^{\text{b-e}}_{\beta,\gamma}\(z_3,z_2;1\)}{\delta \xi^{n}_{\bku}\(z_4\)}
  &=\sum_{\phi,\psi}\int\!dz_5dz_6\frac{\delta
    \Gamma^{\text{b-e}}_{\beta,\gamma}\(z_3,z_2;1\)}{\delta D_{\phi,\psi}\(z_5,z_6\)}
  \frac{\delta D_{\phi,\psi}\(z_5,z_6\)}{\delta\xi^{n}_{\bku}\(z_4\)}
  +\int\!d56\,\frac{\delta \Gamma^{\text{b-e}}_{\beta,\gamma}\(z_3,z_2;1\)}{\delta G\(5,6\)}
  \frac{\delta G\(5,6\)}{\delta\xi^{n}_{\bku}\(z_4\)} \nn\\
  &=\sum_{\phi,\psi}\int\!dz_5dz_6\frac{\delta
    \Gamma^{\text{b-e}}_{\beta,\gamma}\(z_3,z_2;1\)}{\delta D_{\phi,\psi}\(z_5,z_6\)}\sum_{\chi,\lambda}\int\!dz_7dz_8\,
  D_{\phi,\chi}\(z_5,z_7\)\Gamma^{\text{b-b},{n}}_{\chi,\lambda;\bku}\(z_7,z_8;z_4\)D_{\lambda,\psi}\(z_8,z_6\)\nn\\
  &+\int\!d5678\,\frac{\delta \Gamma^{\text{b-e}}_{\beta,\gamma}\(z_3,z_2;1\)}{\delta G\(5,6\)}
  G\(5,7\)\Gamma^{\text{e-b},{n}}_{\bku}\(7,8;z_4\) G\(8,6\).\label{eq:P:3:ab}
\end{align}
\end{widetext}
With this ingredient, the theory of interacting fermions and bosons is formally
complete: the self-energies are expressed in terms of propagators and vertex
functions. Note that we do not have yet determining equations for higher-order bosonic
propagators and for the electron density response functions. For the former, one would
have to study the equation of motion for $\ww{Q}^n_{\bnu}$ which, is rather
compicated. Therefore, in Sec.~\ref{sec:responses} we use again the method of functional
derivatives to recast $\chi$ and $D^{m,n}_{\bmu,\bnu}$ in terms of the simplest propagators
$G$ and $D$.

The vertex funcions are related by the generalized Bethe-Salpeter equation which
retains a surprisingly simple structure pertinent to the pure electronic case. The
relation between bare and dressed vertex functions is a nontrivial point in the theory of
electron-phonon interactions (see Sec.~V.A of
Giustino~\cite{giustino_electron-phonon_2017}). In the case of linear electron-phonon
interactions the vertex is renormalized solely due to the electron-electron interactions
(e.g. Fig.~2 of Leeuwen~\cite{van_leeuwen_first-principles_2004}). In the nonlinear case
considered here, the four vertex functions inevitably arise from a single electron-boson
vertex, $V_{\underline\nu}^{n}(\mathbf{x})$ . At a marked difference with these simpler theories, there are now four ways to
renormalize the bare vertex. In the next Sec.~\ref{sec_lowest} we consider what form the
electron and the boson propagators take when the lowest-order approximations
(Eqs.~\ref{eq:J0}) are adopted for the vertex functions.
%
%
%
%
\section{Lowest--order approximations for the bosonic and electronic self-energies\label{sec_lowest}}
\begin{figure}[]
  {\centering \includegraphics[]{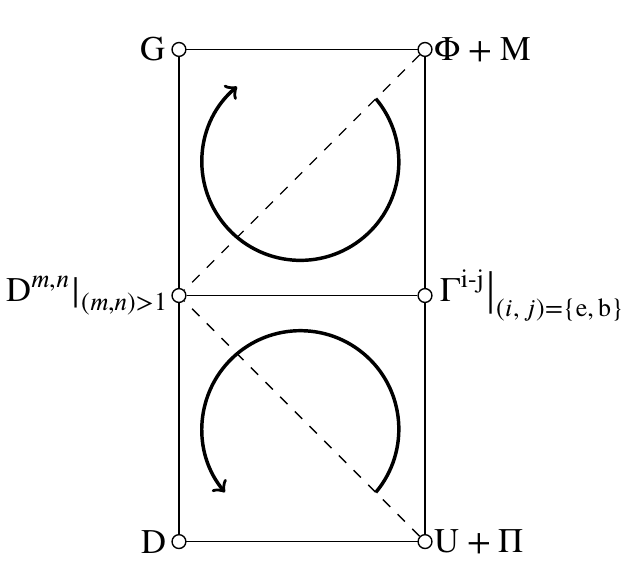}}
  \caption{Schematic representation of the self-consistent cycle involving the different
    components of the generalized Hedin's equations.  The dashed lines correspond to the
    generalized $GW$ approximation where the vertex functions $\Gamma$ are taken to their
    lowest order approximation and $D^{m,n}\approx \left. D^{m,n}\right|_0$}.
  \label{fig:HEDINS}
\end{figure}

The solution of the Dyson equations for fermions and bosons are considerably more involved
than in the case of linear electron--boson coupling. The equations have two level of
internal consistency that we schematically represent in Fig.\,\ref{fig:HEDINS}.

Let us take the electronic case as an example. The Dyson equation is itself nonlinear. For
a given approximation for $M$ the Dyson equation must be solved and the new $G$ plugged in
$M$ for a new solution. This process must be continued up to when self-consistency is
reached. Besides this internal consistency the mass operator depends on the vertex
function $\overline{\Gamma}^\text{e-b}$ and on the multi--boson propagators $D^{n,m}$.
The usual approach to cut this self-consistent loop is based on approximating the vertexes
to their lowest order and to take the independent boson approximation (IBA) for
$D^{n,m}$. A similar procedure can be applied in the bosonic case.

It is interesting to note that, at variance with the purely electronic case, the zeroth
order bosonic vertex functions are still dependent on $D^{n,m}$ through the
$\langle\ww{Q}^{n}_{\bnu}\rangle$ terms appearing in
Eqs.~(\ref{eq:G0.be},\,\ref{eq:G0.bb}). This dependence is resolved in the
self--consistent loop of \fig{fig:HEDINS} by simply looking at the
$\langle\ww{Q}^{n}_{\bnu}\rangle$ as contractions of bosonic response function. Therefore,
for the zeroth order vertexes will use the IBA, $\langle\ww{Q}^{n}_{\bnu}\rangle\approx
\left.D^{n-1,1}_{\(\nu_1\dots\nu_{n-1}\),\nu_n}\(z,z^+\)\right|_0$.

%
%
%
%
%
\subsection{Electrons: the generalized Fan approximation}~\label{sec_Fan}
\begin{figure}[H]
  {\centering
  \includegraphics[]{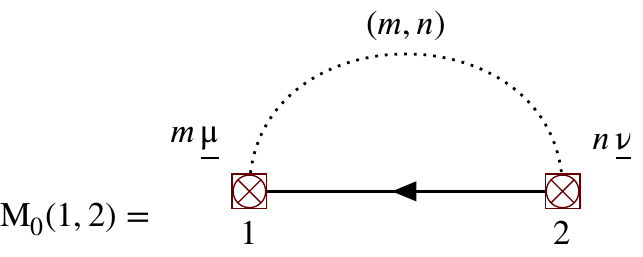}}
  \caption{The lowest--order fermionic self--energy.}
\label{fig:lowest_Pi_f}  
\end{figure}
By using the zeroth--order $\overline\Gamma^\text{b-e}$ vertex function,
\eq{eq:zero_vertex} in the mass operator expression, \eq{eq:Sigma:Gen}, allows to
introduce a generalization of the Fan
approximation~\cite{giustino_electron-phonon_2017,Cardona2005b}.  Indeed we get $M\(1,2\)
\approx M_0\(1,2\)$ with
\begin{multline}
  M_0\(1,2\)= i \sum_{n,\, \bnu} \sum_{m\,\bmu}V_{\bnu}^{n}\(\xx_1\) V_{\bmu}^{m}\(\xx_2\) \\
  \times G\(1,2\) \left. D^{m,n}_{\bmu,\bnu}\(z_2,z_1\)\right|_0.
\label{eq:fan}
\end{multline}
\eq{eq:fan} represents the generalization of the usual Fan approximation which is known
only in the linear coupling case (corresponding to $m=n=1$).  Its diagrammatic form is
shown in \fig{fig:lowest_Pi_f}.

In \eq{eq:fan} $\left. D^{n,m} \right|_0$ is the zeroth--order approximation for the
bosonic propagator which can be recast as a functional of noninteracting bosonic
propagators, as described in Sec.\ref{sec:IP_bosons} for some specific cases.
%
%
%
%
%
\subsection{Bosons: a generalized polarization self-energy}~\label{sec_bose_Fan}
As sketched in \fig{fig:HEDINS}, the lowest order approximation for the bosonic
self-energy is obtained by using the zeroth--order generalized vertex functions,
\eq{eq:J0}, and the IBA\,($D^{n,m}\approx \left. D^{n,m} \right|_0$) and
  IPA\,($\chi\approx\chi_0$) for for bosons and electrons, respectively.

\begin{widetext}
These approximation must be used in \eq{eom:M_1_4}, \eq{eq:U_2p}, \eq{eq:P:1}, \eq{eq:U:1},
\eq{eq:P:2} and \eq{eq:P:3}.  Eq.(\ref{eq:P:3}b) and Eq.(\ref{eq:P:3}c) need not be
considered because it contains variations of other vertex functions.
\begin{figure}[t!]
  {\centering \includegraphics[width=\textwidth]{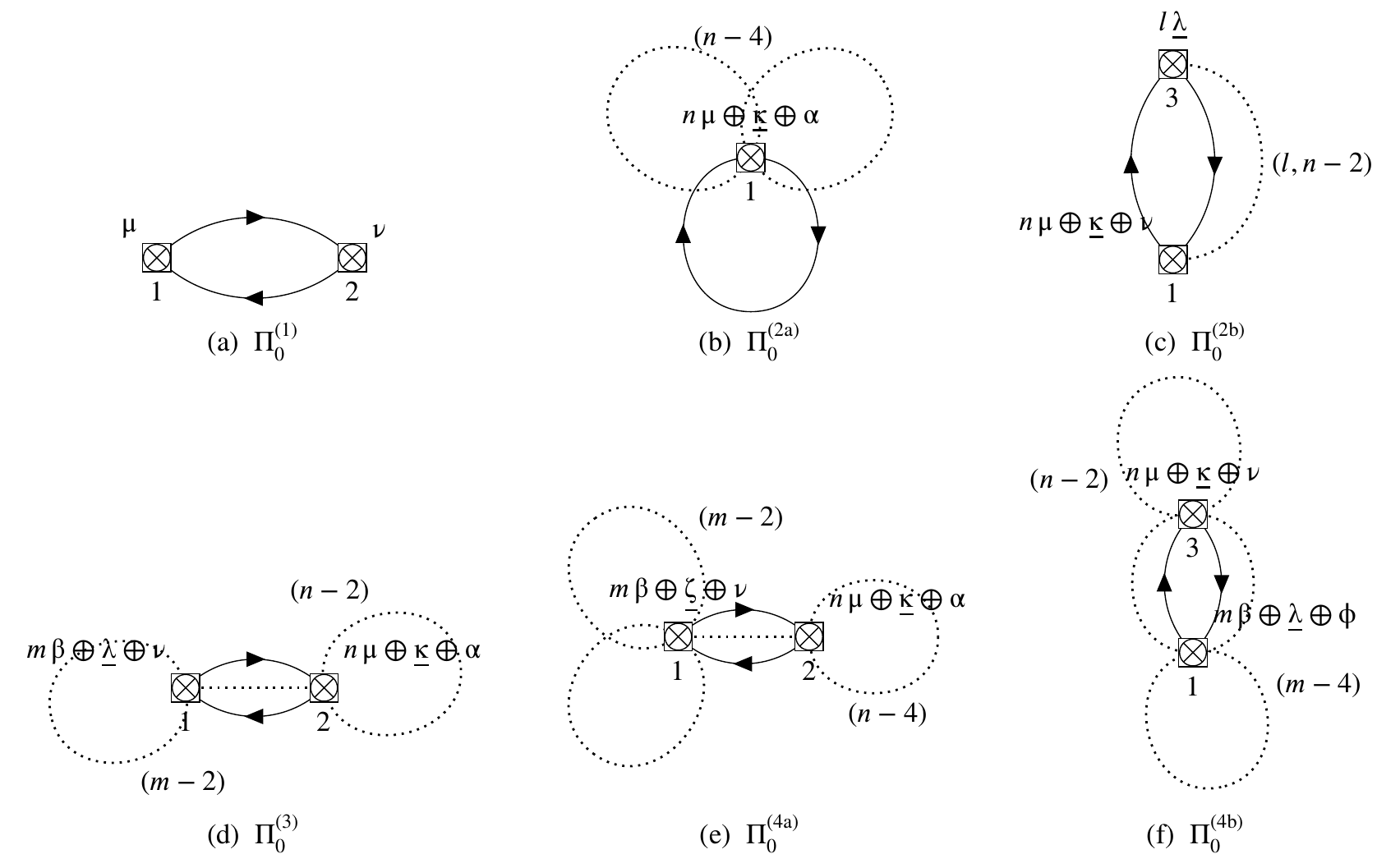}}
  \caption{Lowest--order approximation for the bosonic mass operators,
    $\left.\Pi_{\mu,\nu}\(z_1,z_2\)\right|_0$.\label{fig:P:zero-order}}
\end{figure}
In total we obtain six terms:
\begin{subequations}
  \label{eq:PI_zero}
\begin{align}
  \left.\Pi^{(1)}_{\mu,\nu}\(z_1,z_2\)\right|_0 = \int\!d\xx_1\,d\xx_2\,
  V^{1}_{\mu}\(\xx_1\)G\(1,2\) G\(2,1\)  V^{1}_{\nu}\(\xx_2\),
\end{align}
\begin{align}
\left.\Pi^{\(2a\)}_{\mu,\nu}\(z_1\)\right|_0 =
  \sum_{n \geqslant 2,\,\bku,\ga} n \langle\hat\gamma^n_{\mu\oplus\bku\oplus\alpha}\(z_1\)\rangle
  \sum_{\gb} D_{\alpha,\beta}\(z_1,z_1\)\left.\Gamma^{\text{b-b},n-2}_{\beta,\nu;\bku}\(z_1\)\right|_0,
\end{align}
\begin{align}
  \left.\Pi^{\(2b\)}_{\mu,\nu}\(z_1\)\right|_0 =\sum_{n,\,\bku} n \, \int\! d\xx_1\,
  V^n_{\mu\oplus\bku\oplus\nu}\(\xx_1\)
  \sum_{l,\,\bgl} \int\!d3\,G\(1,3\) V^l_{\bgl}\(\xx_3\) \left. D^{l,n-2}_{\bgl,\bku}\(z_3,z_1\)\right|_0 G\(3,1^+\),
\end{align}
\begin{align}
  \left.\Pi^{\(3\)}_{\mu,\nu}\(z_1,z_2\)\right|_0 =i\sum_{
    \substack{n,m,\,\bku  \\\alpha,\gb,\bgl }} \(n m\)\, \int\! d\xx_1\,
  V^n_{\mu\oplus\bku\oplus\alpha}\(\xx_1\)
  \Langle\ww{Q}^{n-2}_{\bku}\Rangle \int\!d\xx_2\, D_{\alpha,\beta}\(z_1,z_2\)
  V^m_{\beta\oplus\bgl\oplus\nu}\(\xx_2\)
  \Langle\ww{Q}^{m-2}_{\bgl}\Rangle\chi_0\(2,1\),
\end{align}
\begin{multline}
\left.\Pi^{\(4a\)}_{\mu,\nu}\(z_1,z_2\)\right|_0 =
-\sum_{n, \bku, \alpha} n \, \int\! d\xx_1\,V^n_{\mu\oplus\bku\oplus\alpha}\(\xx_1\)
 \sum_{\beta,\phi,\psi} \int\! dz_3 D_{\alpha,\phi}\(z_1,z_1\)
 \left.\Gamma^{\text{b-b},n-2}_{\phi,\psi;\bku}\(z_1\)\right|_0 D_{\psi,\beta}\(z_1,z_3\)
\\\times\sum_{m,\,\bgz,\beta} m \int\!d\xx_2\,V^m_{\beta\oplus\bgz\oplus\nu}\(\xx_2\)
\Langle \ww{Q}_{\bgz}^{m-2}\Rangle D_{\psi,\beta}\(z_1,z_2\)\chi_0\(2,1\),
\end{multline}
\begin{multline}
\left.\Pi^{\(4b\)}_{\mu,\nu}\(z_1\)\right|_0=-\sum_{n,\bku,\beta,\phi,\psi,\alpha}
n\, \int\! d\xx_1\,V^n_{\mu\oplus\bku\oplus\ga}\(\xx_1\) 
\left.\Gamma^{\text{b-b},n-2}_{\psi,\nu;\bku}\(z_1\)\right|_0\\
\times\sum_{m,\,\bgl} m \int\!d\xx_3dz_3\,V^m_{\beta\oplus\bgl\oplus\phi}\(\xx_3\)
\Langle \ww{Q}_{\bgl}^{m-2}\Rangle D_{\alpha,\beta}\(z_1,z_3\)D_{\phi,\psi}\(z_3,z_1\) \chi_0\(3,1\).
\end{multline}
\end{subequations}
These equations are depicted diagrammatically in Fig.~\ref{fig:P:zero-order}.
\end{widetext}
%
%
%
%
\section{Response functions}\label{sec:responses}
The electronic and bosonic self--energies are written, also, in terms of the response
functions, $\chi\(1,2\)$ and $D^{m,n}_{\bmu,\bnu}\(z_1,z_2\)$ with $n>1$ or $m>1$. These
response functions are more involved to calculate compared to the single--body case.
Indeed, in the purely electronic case, the single electronic GF satisfies the Dyson
equation, while the two--bodies GF solves a more complicated Bethe--Salpeter
equation~\cite{onida_electronic_2002}. This is the contracted form of the equation of
motion for the electronic vertex.

However, when the electronic and bosonic
degrees of freedom are considered on equal footing as in Sec.\ref{sec_boson_vertexes},
the four vertex functions are mutually connected via a matrix integro-differential,
\eq{eq:gen:BSE}\,---\,the generalized Bethe-Salpeter equation.

In the following sections our goal is investigate the form which take the electronic and
the bosonic response functions as a consequence of the GBSE. In addition, thanks to the
power of the Schwinger technique of functional derivatives, we will rewrite the equation
of motion for the response function in terms of single fermion and single boson
self-energies.

We have two aspects that complicate enormously the goal of this section: (i) the
electronic and bosonic response functions are mutually dependent, (ii) the $D$ may contain
an arbitrary pair of incoming and outgoing bosonic lines, $\(n,m\)$.

%
%
%
%
%
\subsection{Electronic response}\label{sec:electronic_response}
The electronic response, Eq.~\eqref{eq:e_resp.1}, can be rewritten in terms of the purely
electronic vertex, $\Gamma^\text{e-e}$ by means of the usual chain rule and connecting $\gr$ to
the trace of $G$:
\begin{multline}
  \chi(1,2) =i \int\!d34\, G\(1,3\)\frac{\delta G^{-1}\(3,4\)}{\delta \eta\(2\)}G\(4,1^+\)\\
  =i \int\!d34\, G\(1,3\) G\(4,1^+\) \Gamma^\text{e-e}\(3,4;2\).
 \label{eq:e_resp.2}
\end{multline}
From \eq{eq:gen:BSE} we do know that the equation of motion for $ \Gamma^\text{e-e}$
corresponds to the $\text{e-e}$ channel of GBSE. In practice this means that, at variance
with the purely electronic case, {\em it is not possible to write the equation of motion
for the response function solely in terms of $\chi$}. Indeed, $\chi$ will 
depend, in general, on $D^{n,m}$ and, also, on the two mixed generalized response functions
obtained by contracting $\Gamma^{\text{b-e}}$ and $\Gamma^{\text{e-b}}$ with bosonic and
fermionic operators.

An alternative path, that we follow here, is to find an explicit form of
$\Gamma^\text{e-e}$ and use \eq{eq:e_resp.2} to obtain $\chi$. From \eq{eq:gen:BSE} we
know that
\begin{widetext}
\begin{multline}
  \Gamma^{\text{e-e}}\(3,4;2\)= 
  \Gamma^{\text{e-e}}_0\(3,4,2\)+ \int\,d5678\, K^{\text{e-e}}\(3,6;4,5\) G\(5,7\) G\(8,6\)
  \Gamma^{\text{e-e}}\(7,8;2\)\\
  +\sum_{\psi\xi\phi\eta}\int\!d56\int\!d z_7 z_8\, K^{\text{e-b}}_{\psi,\phi}\(3,z_6;4,z_5\)
  D_{\phi,\xi}\(z_5,z_7\) D_{\eta,\psi}\(z_8,z_6\)\Gamma^{\text{b-e}}_{\xi,\eta}\(z_7,z_8;2\).
  \label{eq:e_resp.4}
\end{multline}
The first two terms in Eq.\,\eqref{eq:e_resp.4} represent a generalization of the usual
Bethe--Salpeter equation, widely used in the context of optical
absorption~\cite{onida_electronic_2002}, to the case of an arbitrary number of bosons that
mediate the electron--hole interaction. The second term, instead, is new and represents a
boson--mediated electron--hole propagation. The electron--hole pair annihilates producing
a number of bosons, which are subsequently scattered giving rise to a particle-hole pair.
\end{widetext}

In order to visualize this important modifications we consider the case where $M$ is
approximated with the generalized Fan form, \eq{eq:fan}, to evaluate $K^{\text{e-e}}$ and
$K^{\text{e-b}}$:
\begin{multline}
  K^{\text{e-e}}\(3,5;4,6\)\approx K_0^{\text{e-e}}\(3,4\) \gd\(3,5\) \gd\(4,6\) \\
  =  i \sum_{n m} \sum_{ \bmu \bnu} V_{\bnu}^{n}\(\xx_3\) V_{\bmu}^{m}\(\xx_4\) 
 \left.D^{m,n}_{\bmu,\bnu}\(z_4,z_3\)\right|_0,
  \label{eq:K_ee_fan}
\end{multline}
and
\begin{multline}
  K^{\text{e-b}}_{\phi,\psi}\(3,z_5;4,z_6\)\\\approx
  \left. K_{\phi,\psi}^{\text{e-b}}\(z_3;z_4\)\right|_0 \gd\(z_3-z_6\)\gd\(z_4-z_5\) \\
  = i G\(3,4\)\sum_{n m} \sum_{\bku\,\bgl}
  \int d \xx_3 \xx_4 V_{\bku\oplus\phi}^{n}\(\xx_3\)  V_{\bgl\oplus\psi}^{m}\(\xx_4\) \\
 \left.D^{m-1,n-1}_{\bgl,\bku}\(z_4,z_3\)\right|_0.
 \label{eq:K_eb_fan}
\end{multline}
We can now use Feynman diagrams to make the different contributions to $\chi$ more
transparent. Let us consider the specific case where we use
$\Gamma^{\text{e-e}}\(6,7;3\)\approx \Gamma_0^{\text{e-e}}\(6,7;3\)$ and
$\Gamma^{\text{b-e}}\(z_6,z_7;3\)\approx \left.\Gamma^{\text{b-e}}\(z_6,z_7;3\)\right|_0$
in the r.h.s. of \eq{eq:e_resp.4}. If we plug 
Eqs.\eqref{eq:K_ee_fan}, \eqref{eq:K_eb_fan} in \eq{eq:e_resp.4}
and the resulting $\Gamma^{\text{e-e}}$ in
\eq{eq:e_resp.2}, a closed form expression for $\chi$ follows.

\begin{figure}
  {\centering
  \includegraphics[]{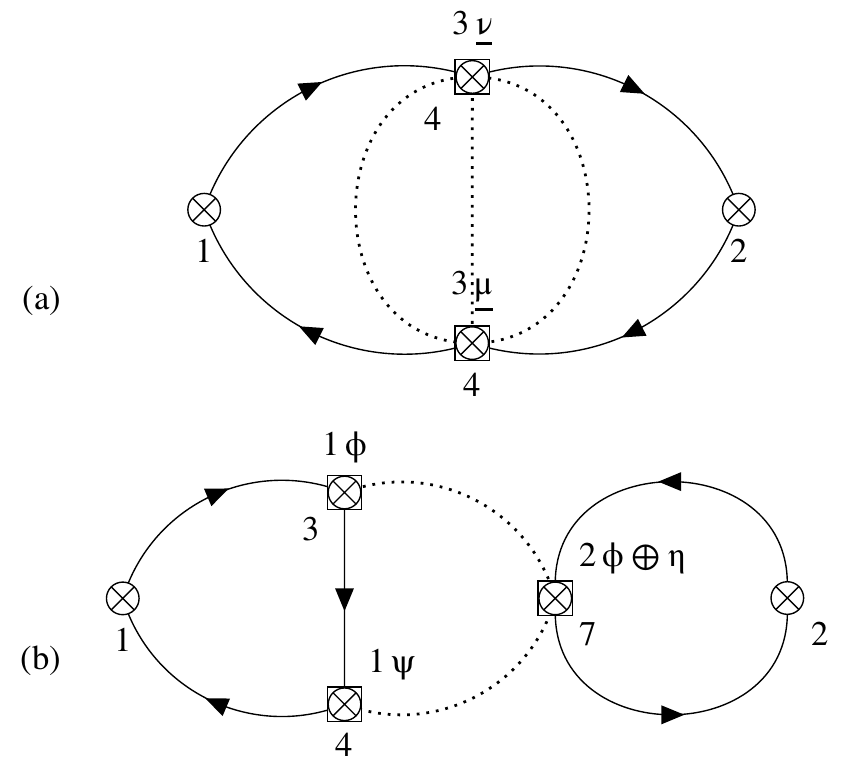}}
  \caption{First order contributios to the electronic response function.  (a) is the
    contribution from $ K_0^{\text{e-e}}$ when $n=2$, while (b) is the contribution
    from $K_0^{\text{e-b}}$ when $n=1$.  Both terms are calculated with $M$ approximated
    with the generalized Fan approximation, \eq{eq:fan}.  Already at this simple
    order of perturbation theory (a) shows the simultaneous electron--hole
    interaction mediated by three bosons.  (b), instead, is totally new and shows how
    the electron--hole dynamics is temporarely transformed in a two bosons dynamics
    already in the linear coupling.
\label{fig:chi_e_first_order}}
\end{figure}

In \fig{fig:chi_e_first_order} we consider two interesting cases of \eq{eq:e_resp.4}: (a)
the contribution from the first integral and $K^{\text{e-e}}$ evaluated with $n=m=3$, (b)
the contribution from the second integral when $n=m=1$ in $K^{\text{e-b}}$ and $n=2$ in
$\Gamma^{\text{b-e}}$.
%
%
%
%
%
\subsection{Bosonic response}\label{sec:resp_bosons}
We start from \eq{eq:D_as_dD}, applied to $D^{n+\Delta n,n}$. Thanks to this equation it
is possible, for a given $n$, to reduce the evaluation of $D^{n+\Delta n,n}$ to the one of
$D^{n,n}$, $D^{\Delta n,n}$ and the functional derivative of $D^{n,n}$. If we assume
$\Delta n \leqslant n$ (the derivation can be easily extended to the case $\Delta n>n$)
\eq{eq:D_as_dD} lowers the order of $n+\Delta n$. If we further apply the same procedure
to $D^{\Delta n,n}=D^{n,\Delta n}=D^{\Delta n+\(n-\Delta n\),\Delta n}$ the initial
problem of evaluating $D^{n+m,n}$ can be cast in an expression which includes only
diagonal response function, of the form $D^{m,m}$ with $m$ an arbitrary integer
$m\leqslant n$.

Let us take as an example the $D^{5,2}\(z_1,z_2\)$ case. From \eq{eq:D_as_dD} it follows
that
\begin{multline}
  D^{5,2}_{\bmu,\bnu}\(z_1,z_2\)= i\frac{\delta}{\delta\xi_{\bku}^{3}\(z_1\)}D^{2,2}_{\bgl,\bnu}\(z_1,z_2\)\\
  +\Langle \ww{Q}_{\bku}^3\(z_1\)\Rangle D^{2,2}_{\bgl,\bnu}\(z_1,z_2\)\\
  +\Langle \ww{Q}^{2}_{\bgl}\(z_1\)\Rangle D^{3,2}_{\bku,\bnu}\(z_1,z_2\),
\end{multline}
with $\bmu=\bku\oplus\bgl$. We can now apply again \eq{eq:D_as_dD} on
$D^{3,2}_{\bku,\bnu}\(z_1,z_2\)$. It follows that
\begin{multline}
  D^{3,2}_{\bku,\bnu}\(z_1,z_2\)= i\frac{\delta}{\delta\xi_{\gb}^{1}\(z_1\)}D^{2,2}_{\bga,\bnu}\(z_1,z_2\)\\
  +\Langle \ww{Q}^{2}_{\bga}\(z_1\)\Rangle D^{1,2}_{\gb,\bnu}\(z_1,z_2\),
\end{multline}
with $\bku=\gb\oplus \bga$. A last application of \eq{eq:D_as_dD} finally gives
\begin{align}
  D^{1,2}_{\gb,\bnu}\(z_1,z_2\)=D^{2,1}_{\bnu,\gb}\(z_2,z_1\)= i\frac{\delta}{\delta\xi_{\nu_2}^{1}\(z_2\)}
  D_{\nu_1,\gb}\(z_2,z_1\).
\end{align}
We have finally reduced $D^{5,2}$ to an explicit functional of only diagonal response
functions and their derivatives: $D^{5,2}=F\[D,D^{2,2},\frac{\gd D}{\gd \xi^1}, \frac{\gd
  D^{2,2}}{\gd \xi^1},\frac{\gd D^{2,2}}{\gd \xi^3}\]$.  From this simple example it
follows that it is enough to study diagonal bosonic response functions and their
functional derivatives in order to calculate any non--diagonal response functions.

In the following we discuss the IBA and give as an example the case of $D^{2,2}$ and
$D^{3,3}$.
\subsubsection{The independent bosons approximation}\label{sec:IP_bosons}
The limit of independent bosons is instructive to understand the actual number of diagrams
that can be expected at any level of the perturbative expansion.  In order to evalute this
number in the IBA we observe that:
\begin{multline}
  \left. D^{m,n}_{\bmu,\bnu}\(z_1,z_2\)\right|_0=
  -i\Langle\mathcal{T}\wl
    \Delta\ww{Q}_{\mu_1}\(z_1\)\dots\Delta\ww{Q}_{\mu_m}\(z_1\)\capo
    \Delta\ww{Q}_{\nu_1}\(z_2\)\dots\Delta\ww{Q}_{\nu_n}\(z_2\)\wr\Rangle_0.
\label{eq:D0}
\end{multline}
with $\langle\cdots\rangle_0$ the thermal average corresponding to the free--bosons
Hamiltonian.  $\left.D^{m,n}_{\bmu,\bnu}\right|_0$ reduces to the sum of all possible
contractions of two bosonic operators.  From simple combinatorics arguments we know that
the number of possible ordered pairs of two operators out of a product of $n\ge 2$ is
given by the number of the so-called chord diagrams~\cite{pavlyukh_analytic_2007}
\begin{align}
  N_n=\left\{
  \begin{array}{cr}
    (n-1)!!&\text{even $n$,}\\
    0      &\text{odd $n$}.
    \end{array}\right.
\end{align}
By doing simple diagrammatic expansion we see, indeed, that
$\left. D^{2,2}_{\bmu,\bnu}\(z_1,z_2\)\right|_0$ produces a total of $N_4=3$ terms. One of
them is disconnected and corresponds to the complete contractions of the two terms
$\Langle\Delta\ww{Q}_{\mu_1}\(z_1\)\Delta\ww{Q}_{\mu_2}\(z_1\)\Rangle_0$ and
$\Langle\Delta\ww{Q}_{\nu_1}\(z_2\)\Delta\ww{Q}_{\nu_2}\(z_2\)\Rangle_0$.

In the $\left. D^{3,3}_{\bmu,\bnu}\(z_1,z_2\)\right|_0$ case, instead, all
contractions are connected because there is always at least one contraction with
different time-arguments. This means that we have in total $N_6=15$ terms.  The
explicit form of $\left.  D^{3,3}_{\bmu,\bnu}\(z_1,z_2\)\right|_0$ will be given in the
Sec.\ref{sec:D33}.

We can therefore generally state that
$\left. D^{n,m}_{\bmu,\bnu}\(z_1,z_2\)\right|_0$ is composed of $N_{n+m}-N_nN_m$
connected diagrams.

This simple combinatorics discussion allows us to derive some general rule on the strength
of the $n$th order of the perturbative expansion. As it is clear from the derivation done
in the precedent sections at any order of the perturbative expansion, a
$D^{m,n}_{\bmu,\bnu}$ appears multiplied by $V_{\bnu}^{n} V_{\bmu}^{m}$. These potentials
include a $1/\(n!m!\)$ prefactor.

Overall, we can deduce that the $\(n,m\)$ order in the bosonic propagator will be weighted
with a $N_{n+m}/\(n!m!\)$ prefactor. When $n$ increases this term decays fast enough to
make the overall expansion controllable.
\subsubsection{The two--bosons case}\label{sec:two_bosons}
\begin{figure}[t!]
  {\centering
  \includegraphics[]{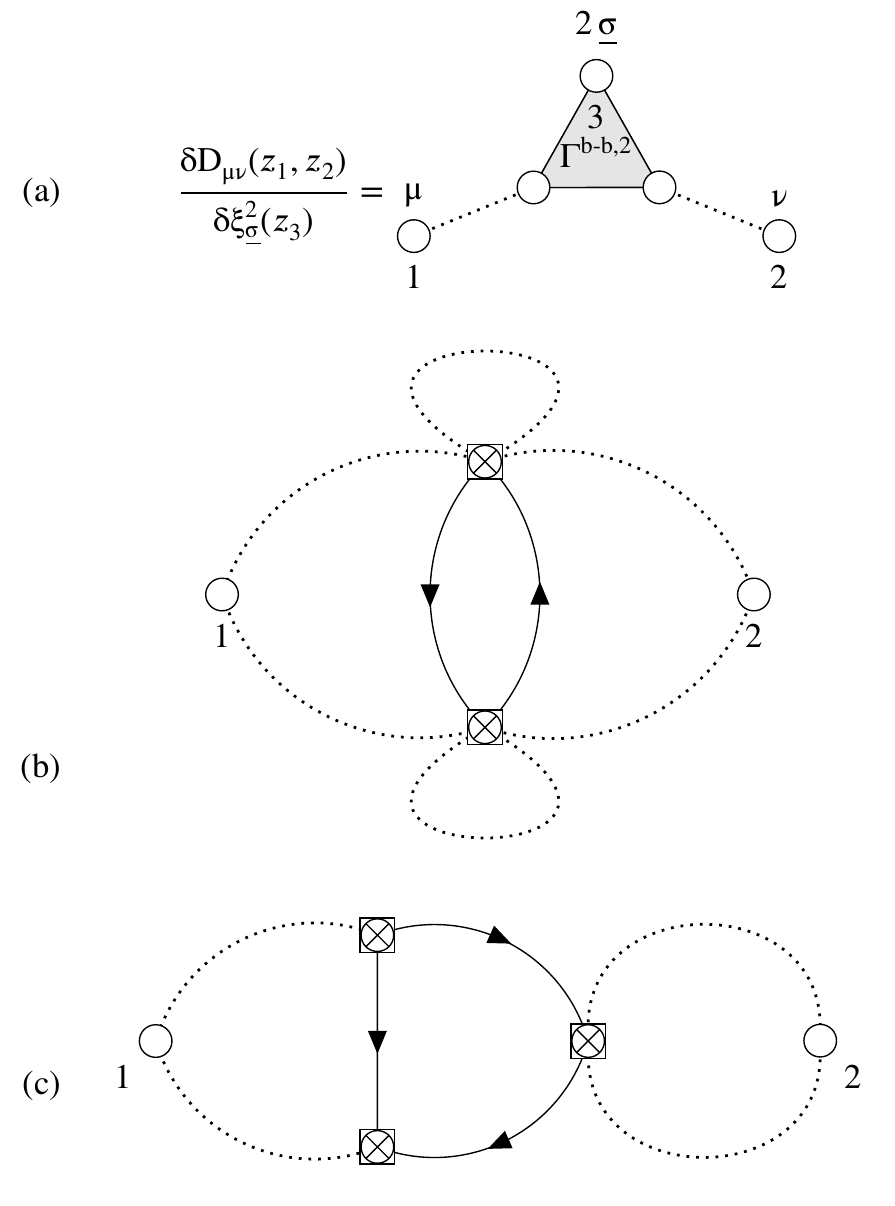}}
  \caption{(a) Digrammatic representation of the first order functional derivative of $D$,
    \eq{eq:b_resp.1}.  (b) and (c) represent two terms contributing to $ D^{2,2}$ which
    show the connection between the single--boson self--energy and the bosonic response
    function. Indeed the diagram\,(b) comes from the scattering term in $K^{\text{b-b}}$
    due to $\frac{\gd \left.\Pi^{\(3\)}\right|_0}{\gd D}$.  Similarly diagram\,(c) is
    induced by the contribution of $\frac{\gd \left.\Pi^{\(1\)}\right|_0}{\gd G}$ to
    $K^{\text{b-e}}$. Both terms are treated, in (b) and (c), at the first order in the
    generalized Bethe-Salpeter equation, \eq{eq:gen:BSE}.
\label{fig:D22}}
\end{figure}
The case of $D^{2,2}$ can be easily worked out following an approach similar to what has
been used in Sec.\ref{sec:electronic_response}.  From \eq{eq:dvdxi} we know that
\begin{align}
D^{2,2}_{\bmu,\bnu}\(z_1,z_2\)=\frac{\delta \langle \ww Q_{\bmu}^2\(z_1\)\rangle}{\delta\xi_{\bnu}^{2}\(z_2\)}.
\label{eq:b_resp.0}
\end{align}
At the same time we can rewrite the $\langle \ww Q_{\bmu}^2\(z_1\)\rangle$ in terms of the
single--body GF, using \eq{eq:D_as_dQ_dxi}:
\begin{align}
\langle \ww Q_{\bmu}^2\(z_1\)\rangle= i D_{\mu_1\mu_2}\(z_1,z_1^+\).
\label{eq:b_resp.0a}
\end{align}
By applying the chain rule we get:
\begin{align}
  D^{2,2}_{\bmu,\bnu}\(z_1,z_2\)=\frac{\delta \langle \ww Q^2_{\bmu} \(z_1\)\rangle }{\delta \xi^{2}_{\bnu}\(z_2\)}=
  i \frac{\delta D_{\mu_1,\mu_2}\(z_1,z_1^+\)}{\delta \xi^{2}_{\bnu}\(z_2\)}.
\label{eq:b_resp.1}
\end{align}
We can now follow the procedure for the electronic case and connect $D^{2,2}$ to the
$\Gamma^\text{b-b}$ vertex:
\begin{multline}
i \frac{\delta D_{\mu_1,\mu_2}\(z_1,z_1\)}{\delta \xi^{2}_{\bnu}\(z_2\)}=
i \sum_{\bga} \int\, dz_3 dz_4 D_{\mu_1\ga_1}\(z_1,z_3\)\\
\Gamma^\text{b-b,2}_{\ga_1,\ga_2;\bnu}\(z_3,z_4;z_2\)
 D_{\ga_2,\mu_2}\(z_4,z_1\).
\label{eq:b_resp.2}
\end{multline}
\eq{eq:b_resp.2} is represented diagrammatically in \fig{fig:D22}(a).

Eq.\,\eqref{eq:gen:BSE} provides the equation of motion for
$\Gamma^\text{b-b,2}_{\ga_1,\ga_2}$ that, in a similar way to \eq{eq:e_resp.4}, is written
in terms of the pure bosonic\,($b-b$) and the mixed boson--electron\,($b-e$) vertex
functions. This equation involves the kernels $K^{\text{b-b}}$ and $K^{\text{b-e}}$.  We
can now follow the same path of the purely electronic case and use the lowest-order
bosonic self--energy, \eq{eq:PI_zero}, to derive the corresponding expression for the b--b
and b--e kernels and, consequently, of $\Gamma^\text{b-b,2}$. Two representative diagrams
contributing to $D^{2,2}$ are shown in \fig{fig:D22}(b) and \fig{fig:D22}(c).

The IBA for $D^{2,2}$ can be easily evaluated by using the zeroth order expression for
$\Gamma^\text{b-b,2}$. From \eq{eq:v_bb.3} we know that when $n=2$ we have only
$n!/(n-1)!=2$ terms,
\begin{multline}
  \left.\Gamma^{\text{b-b},2}_{\ga_1,\ga_2;\bnu}\(z_1,z_2;z_3\)\right|_0 =\\
  \gd\(z_1-z_2\)\gd\(z_1-z_3\) \[\gd_{\ga_1,\nu_1}\gd_{\ga_2,\nu_2}+ \gd_{\ga_1,\nu_2}\gd_{\ga_2,\nu_1}\],
\label{eq:b_resp.3}
\end{multline}
which gives
\begin{multline}
 \left. D^{2,2}_{\bmu,\bnu}\(z_1,z_2\)\right|_0= i \[ D_{\mu_1,\nu_1}\(z_1,z_2\) D_{\nu_2,\mu_2}\(z_2,z_1\)+\capo 
      D_{\mu_1,\nu_2}\(z_1,z_2\) D_{\nu_2,\mu_1}\(z_2,z_1\)\].
\label{eq:b_resp.4}
\end{multline}
\eq{eq:b_resp.4} coincides with the expression that can be derived by using the
diagrammatic approach.  
\subsubsection{The three--bosons case}\label{sec:D33}

\begin{figure}[t!]
  {\centering
  \includegraphics[width=9cm]{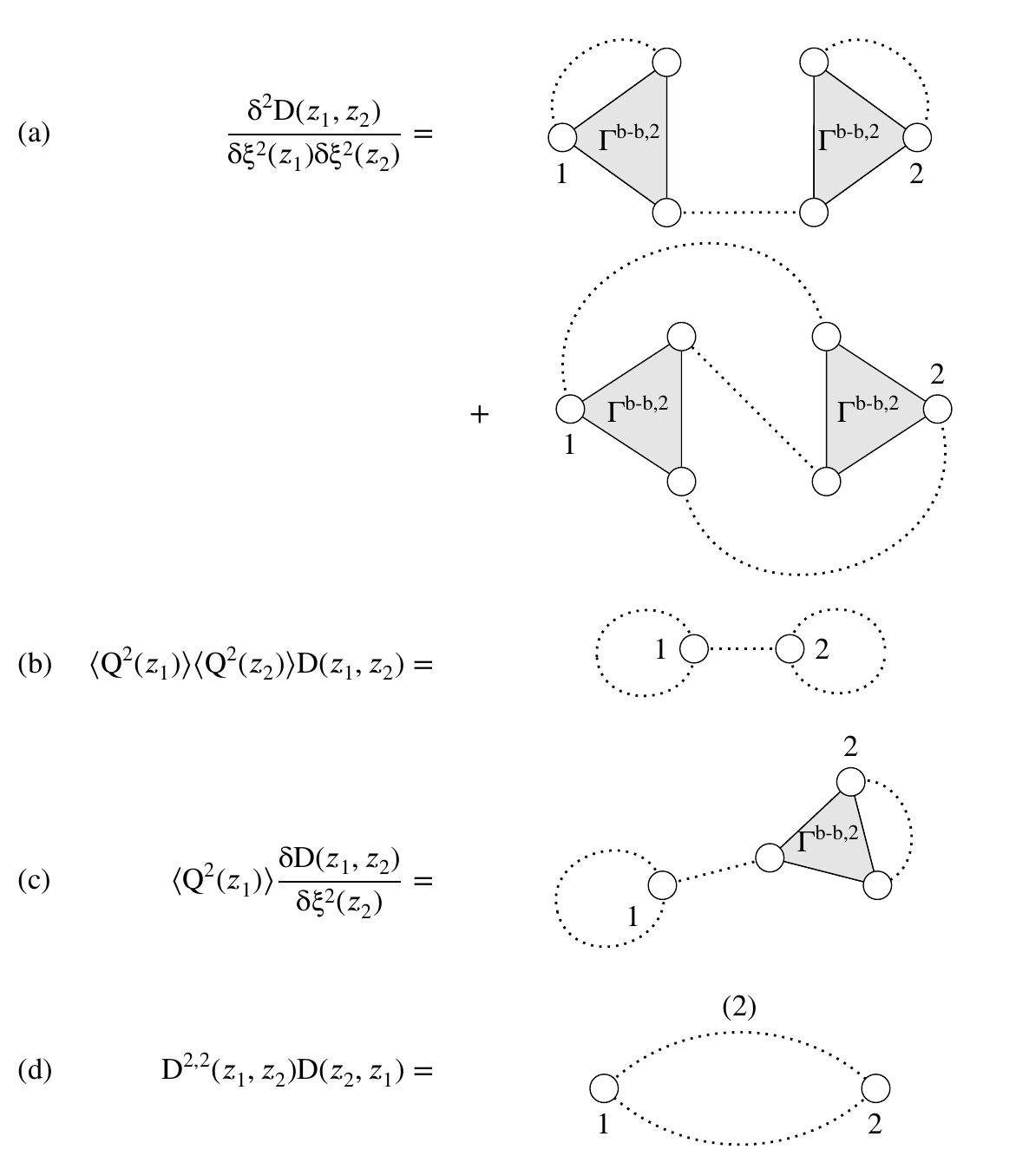}}
  \caption{Diagrammatic representation of the terms in \eq{eq:D33.2} contributing to
    $D^{3,3}$.
    \label{fig12}}
\end{figure}
In the three--bosons case the calculation of $D^{3,3}$ may appear to be prohibitively
complicated. Still, the present scheme allows, via the functional derivative approach to
derive it in an elegant and compact way. We start by applying \eq{eq:D_as_dD} to
$D^{3,3}$:
\begin{align}
  D^{3,3}_{\bmu,\bnu}\(z_1,z_2\)=\[i\frac{\delta }{\delta\xi_{\bgl}^{2}\(z_1\)}
  +\Langle \ww Q_{\bgl}^2\(z_1\)\Rangle\] D^{1,3}_{t,\bnu}\(z_1,z_2\),
\label{eq:D33.0}
\end{align}
with $\bmu\equiv \bgl\oplus t$. 
By using \eq{eq:D_as_dD} again we get that
\begin{multline}
D^{1,3}_{t,\bnu}\(z_1,z_2\)=D^{3,1}_{\bnu,t}\(z_2,z_1\)= \\
\[i\frac{\delta }{\delta\xi_{\bgs}^{2}\(z_2\)}+\Langle \ww Q_{\bgs}^2\(z_2\)\Rangle\]D_{s,t}\(z_2,z_1\),
\label{eq:D33.1}
\end{multline}
where $\bnu\equiv \bgs\oplus s$. Eq.\,\eqref{eq:D33.0} and \eq{eq:D33.1} show that
$D^{3,3}$ is composed of five terms
\begin{multline}
  D^{3,3}_{\bmu,\bnu}\(z_1,z_2\)= \Bigg[D^{2,2}_{\bgl,\bgs}\(z_1,z_2\)
    -\frac{\delta^2}{\delta\xi_{\bgl}^{2}\(z_1\)\xi_{\bgs}^{2}\(z_2\)}\\
    +\Langle \ww Q_{\bgl}^2\(z_1\)\Rangle\Langle \ww Q_{\bgs}^2\(z_2\)\Rangle
    +i\Langle \ww Q_{\bgs}^2\(z_2\)\Rangle\frac{\delta }{\delta\xi_{\bgl}^{2}\(z_1\)}\\
    +i\Langle \ww Q_{\bgl}^2\(z_1\)\Rangle\frac{\delta }{\delta\xi_{\bgs}^{2}\(z_2\)}\Bigg]D_{s,t}\(z_2,z_1\).
\label{eq:D33.2}
\end{multline}
The construction of diagrammatic form of \eq{eq:D33.2} can be done by using a simple
diagrammatic form of the \eq{eq:b_resp.2}, as shown in \fig{fig:D22}(a). This shows that
any of the functional derivatives appearing in \eq{eq:D33.2} can be rewritten in terms of
a second order b--b vertex function. In this way it is possible to rewrite $D^{3,3}$ in
terms of known quantities, as shown in \fig{fig12}.  All diagrams represented in
\fig{fig12} reduce, when $\Gamma^{\text{b-b},2}\approx
\left.\Gamma^{\text{b-b},2}\right|_0$ to the IBA expression for $D^{3,3}$ which is,
indeed, composed of a total of 15 terms.

%
%
%
%
%
\section{Conclusions}\label{sec:concs}
In this work we applied Schwinger's variational derivative technique to calculate the
coupled electronic and bosonic dynamics induced by an electron--boson Hamiltonian with
coupling linearly proportional to the electronic density $\hat{n}(\vec{x})$ and to all
orders in the bosonic displacement $\ww{Q}_\nu$. 

The complex and coupled electronic and bosonic dynamics is formulated in the form of a
system of functional relations between the dressed electronic $G\(1,2\)$, the {\em single
  boson} $D_{\mu,\nu}\(z_1,z_2\)$ propagators and the generalized electronic and bosonic
self--energies, $\Sigma^\text{e}\(1,2\)$ and $\Sigma^\text{b}\(z_1,z_2\)$.

These are expressed as closed functionals of the electron density-density response $\chi$,
the multi--boson response functions $D^{n,m}$, and four different vertex functions:
$\Gamma^{\text{e-e}}$, $\Gamma^{\text{b-e}}$, $\Gamma^{\text{e-b}}$ and
$\Gamma^{\text{b-b}}$.  These vertex functions are shown to have either a mixed
electron--boson character ($\Gamma^{\text{b-e}}$ and $\Gamma^{\text{e-b}}$), or a purely
electronic ($\Gamma^{\text{e-e}}$) and bosonic ($\Gamma^{\text{b-b}}$) character.  The
exact equations of motion for all these quantities are formally derived. Sound and
controlled approximations are also proposed in order to make the calculations feasible.

The present formulation allows us to tackle the very ambitious problem of deriving using
the Schwinger's technique coupled equations of motion for the electronic and bosonic
response functions and provide several interesting conclusions and new concepts.

We extend to the nonlinear e--b interaction known concepts like the Debye--Waller
potential and the Fan approximation. We further extend the Bethe--Salpeter equation to a
$2\times 2$ non--linear system of integro-differential equations for the four vertex
functions. Thanks to this equation we show that there is no simple way to decouple the
electronic and bosonic dynamics. We demonstrate, by using simple diagrammatic examples,
that electrons and bosons can equally well mediate the electron--hole and boson--boson
interaction. The present scheme, indeed, demonstrates a full and deep symmetry between the
electronic and bosonic degrees of freedom.

The final result is an important generalization of the well--known Hedin's equations with
a wealth of potential applications in different areas of condensed matter physics, optics
and chemistry.

\section*{Acknowledgements}
AM acknowledges the funding received from the European Union project MaX {\em Materials
  design at the eXascale} H2020-EINFRA-2015-1, Grant agreement n. 676598 and {\em
  Nanoscience Foundries and Fine Analysis - Europe} H2020-INFRAIA-2014-2015, Grant
agreement n. 654360. Y.P. acknowledges funding of his position by the German Research
Foundation (DFG) Collaborative Research Centre SFB/TRR 173 ``Spin+X''.

\appendix
\section{The mean--field treatment of the electron--electron interaction}\label{mean_field}
In order to describe how we treat the correlation induced by the electron--electron
interaction let us start from the full Hamiltonian
  in the first quantization and make explicit the distinction
between dressed and undressed operators:
\begin{align}
\ww{H}= \ww{H}^0_\text{e}+\ww{H}^0_\text{b}+\ww{H}^0_\text{e--b}+\ww{H}_\text{e--e},
\label{eq:A.1}
\end{align}
with the ${}^0$ superscript indicating bare operators. Indeed the dressing of the
different components of the Hamiltonian (when possible) is a product of the dynamics and
cannot be, {\em a priori}, inserted from the beginning.

In \eq{eq:A.1} we introduced
\begin{align}
\ww{H}_\text{e--e}=\frac{1}{2}\sum_{i\neq j}v\(\xx_i-\xx_j\),
\label{eq:A.2}
\end{align}
with $v$ the bare Coloumb potential. It is well documented in the literature that one of
the effects of $\ww{H}_\text{e--e}$ is to screen itself and all other interactions,
including the e--b one. This has been extensively demonstrated, for example, in
Ref.~\onlinecite{marini_many-body_2015}.

The path we take here is, therefore, to embody $\ww{H}_\text{e--e}$ in a mean--field
correction to $\ww{H}^0_\text{e}$ and, consequently, dressing of $\ww{H}^0_\text{b}$ and
$\ww{H}^0_\text{e--b}$:
\begin{align}
\ww{H}\Rightarrow  \[\ww{H}^0_\text{e}+\ww{V}_\text{mf}\]+\ww{H}_\text{b}+\ww{H}_\text{e--b},
\label{eq:A.3}
\end{align}
with $\ww{H}_\text{e}=\ww{H}^0_\text{e}+\ww{V}_\text{mf}$. \eq{eq:A.3} is the connection
with \eq{eq:1.0}. The specific form of $\ww{V}_\text{mf}$ depends on the physical problem.
An example is to use DFT,
where $\ww{V}_\text{mf}=\ww{V}_\text{Hxc}$ is the Hartree plus the Kohn--Sham
  exchange--correlation potential~\cite{dreizler_density_1990}. In this case also the
dressing of $\ww{H}^0_\text{e--b}$ and $\ww{H}^0_\text{e-e}$ is well--known and widely
documented. In the case of the electron--phonon problem, for example, the self--consistent
dressing of the electron--nuclei interaction is described by the Density--Functional
perturbation theory (DFPT)~\cite{Gonze1995,Stefano2001}.

\section{Connection with the electron--phonon problem}\label{e-p-connect}
A specific physical application of the present theoretical scheme is represented by the
coupled electron--phonon system.  This is a very wide field with a wealth of application
in several branches of physics.

The Hamiltonian of the coupled electron--phonon system is obtained by starting from the
total Hamiltonian of the system, that we divide in its independent bare electronic
$\ww{H}_\text{e}$, nuclear $\ww{H}_\text{n}\(\RR\)$, electron--nucleus\,(e--n)
$\ww{W}_\text{e--n}\(\RR\)$ parts
\begin{align}
\ww{H}\(\RR\)= \ww{H}_\text{e}+\ww{H}_\text{n}\(\RR\)+\ww{W}_\text{e--n}\(\RR\),
\label{eq:B.1}
\end{align}
where $\RR$ is a generic notation representing positions of the nuclei. The notation
used in this paper is the same adopted in Ref.~\onlinecite{marini_many-body_2015}.

In introducing \eq{eq:B.1} it is important to stress that $\ww{H}_\text{n}\(\RR\)$
includes both the kinetic and nuclear--nuclear interaction while
$\ww{W}_\text{e--n}\(\RR\)$ represents the electron--nuclei interaction, whose expansion
in the atomic displacements leads, as well known, to the diagrammatic expansion.
Moreover, in the spirit of Appendix~\ref{mean_field} we have assumed, in \eq{eq:B.1}, to
use DFT to describe the effect of the electron--electron correlation via the well--known
exchange--correlation potential.

We split, now, the generic atomic position operator, $\ww{\RR}_I$, in its reference plus
displacement
\begin{align}
\ww{\RR}_I\equiv \overline{\RR}_{I}\hat{\mathbf{1}}+\Delta \ww{\RR}_{I}.
\label{eq:B.2}
\end{align}
The Cartesian components of $\Delta \ww{\RR}_{I}$ play the role of the bosonic coordinate
operators, $Q_{\nu}$. We can, indeed, write that
\begin{align}
\Delta\hat{\RR}_{I}=
\sum_{\nu} \(N M_I \Omega_\nu \)^{-1/2} \bs{\eta}\(\nu|I\) \ww{Q}_{\nu},
\label{eq:B.3}
\end{align}
with $N$ the number of atoms in the system, $M_I$ the mass of atom $I$, $\bs{\eta}$ is the
phonon mode polarization vector.  We assume here, for simplicity, a finite system that can
be generalized to an periodic solid using periodic boundary conditions.

Our initial system is, therefore, characterized by a set of dressed, electronic and
bosonic single--particle states with energies $\{\mE_i\}$ and frequencies $\{\gO_\nu\}$.
We have in total $3N$ bosonic coordinates.

We have now all ingredients to expand the $\ww{W}_\text{e--n}\(\RR\)$ in terms of
$\hat\psi\(\xx\)$ and $\ww{Q}_\nu$. Indeed we can, formally, write that
\begin{multline}
\ww{W}_\text{e--n}\(\RR\)=
\sum_n \ww{W}^{\(n\)}_\text{e-n}\(\RR\)=\\=
\sum_n \sum_{\bs{\nu}} \int\!d\xx\,\hat \psi^\dagger\(\xx\) V^{\(n\)}_{\bs{\nu}}\(\xx\) \hat\psi\(\xx\)\ww{Q}^n_{\bs{\nu}},
\label{eq:B.4}
\end{multline}
with 

\begin{align}
  V^{\(n\)}_{\bs{\nu}}\(\xx\)=
  \(\prod_{i=1}^n \partial_{\nu_i}\)_{eq}
 V_\text{scf}\(\xx-\RR\).
\label{eq:B.5}
\end{align}
In \eq{eq:B.5} $V_\text{scf}$ is the dressed DFPT electron--nuclei potential and the
derivative is taken at the equilibrium position $\RR=\overline{\RR}$.

\section{Proof of \eq{eq:dP} }\label{app_eq_dP}
The equation of motion for $\ww P$ can be derived by using some care. Indeed Eq.(\ref{eq:comm}b) implies that
\begin{multline}
  \[\ww P_\nu\(z_1\),\ww Q^{m}_{\bga}\(z_1\)\]_-=\[\ww P_\nu\(z_1\),\prod_{i=1}^{m} \ww Q_{\alpha_i}\(z_1\)\]_-\\
  =\(-i\)\sum_{j=1}^{m} \gd_{\nu,\alpha_j} \prod_{i\neq j,i=1}^{m} \ww Q_{\alpha_i}\(z_1\).
 \label{eq:P_comm}
\end{multline}
If we now plug \eq{eq:P_comm} into the $\[\ww P_\nu\(z_1\),\ww H\(z_1\)\]_-$ commutator we get
\begin{multline}
\(-i\)\[\ww P_\nu\(z_1\),\ww H\(z_1\)\]_-= \frac{d}{dz_1}\ww P_\nu\(z_1\)=
-\Omega_\nu \ww Q_\nu\(z_1\)\\
-\sum_{m,\bga}  \sum_{j=1}^{m} \gd_{\nu,\alpha_j} 
 \ww \gc_{\bga}\(z_1\) 
\prod_{i\neq j,i=1}^{m}  \ww Q_{\alpha_i}\(z_1\).
 \label{eq:P_comm_1}
\end{multline}
Now we reorder the components of $\bga$ vector ($\gc$ is a fully symmetric tensor) so that
\begin{align}
 \gd_{\nu,\alpha_j} \gc_{\bga}\(z_1\)=  \gc_{\alpha_1,\dots,\alpha_{j-1},\nu,\alpha_{j+1},\dots,\alpha_m}\(z_1\)
 = \gc_{\alpha_1,\dots,\alpha_{m-1},\nu}\(z_1\).
 \label{eq:P_comm_2}
\end{align}
We now rename $\bga$ by introducing the $m-1$ dimensional vector $\bmu\equiv\(\alpha_1,\dots,\alpha_{m-1}\)$. Thanks to \eq{eq:P_comm_2} we have that
\begin{align}
\prod_{i\neq j,i=1}^{m} \ww Q_{\alpha_i}\(z_1\)=\ww Q^{m-1}_{\bmu}\(z_1\),
\end{align}
and  we finally get
\begin{align}
\frac{d}{dz_1}\ww P_\nu\(z_1\)=-\Omega_\nu \ww Q_\nu\(z_1\)-\sum_{m,\bmu} m \,\h{\gamma}^{m}_{\bmu\oplus\nu}\(z_1\)\ww Q^{m-1}_{\bmu}\(z_1\).
 \label{eq:P_comm_final}
\end{align}

\section{Proof of \eq{eq:D_as_dD}}\label{app_D_as_dD_proof}
We start by observing that
\begin{multline}
\frac{i\gd}{\gd \xi_{\bga}^k\(z_1\)} D^{m-k,n}_{\bgb,\bnu}\(z_1,z_2\)=\\
\frac{\gd}{\gd \xi_{\bga}^k\(z_1\)}
\[\Langle {\mathcal T}\ww{Q}^{m-k}_{\bgb}\(z_1\)\ww{Q}^n_{\bnu}\(z_2\)\Rangle\capo-\Langle \ww{Q}^{m-k}_{\bgb}\(z_1\)\Rangle\Langle
\ww{Q}^{n}_{\bnu}\(z_2\)\Rangle\].
\label{eq:F.1}
\end{multline}
We start by expanding the three terms resulting from the functional derivative of the three components of $D$: 
\begin{multline}
\frac{\gd}{\gd \xi_{\bga}^k\(z_1\)}\Langle {\mathcal T}\ww{Q}^{m-k}_{\bgb}\(z_1\)\ww{Q}^n_{\bnu}\(z_2\)\Rangle=\\
\Langle {\mathcal T}\ww{Q}^{m-k}_{\bgb}\(z_1\)\ww{Q}^n_{\bnu}\(z_2\)\ww{Q}^k_{\bga}\(z_1\) \Rangle\\-\Langle {\mathcal
T}\ww{Q}^{m-k}_{\bgb}\(z_1\)\ww{Q}^n_{\bnu}\(z_2\)\Ra\La\ww{Q}^k_{\bga}\(z_1\) \Rangle.
\end{multline}
The second and third term are due to the derivative of the two single displacement operator averages:
\begin{multline}
\frac{\gd}{\gd \xi_{\bga}^k\(z_1\)}\Langle \ww{Q}^{m-k}_{\bgb}\(z_1\)\Rangle=
\Langle \ww{Q}^{m-k}_{\bgb}\(z_1\)\ww{Q}^k_{\bga}\(z_1\) \Rangle\\-\Langle  \ww{Q}^{m-k}_{\bgb}\(z_1\)\Ra\La\ww{Q}^k_{\bga}\(z_1\) \Rangle.
\end{multline}
and
\begin{multline}
\frac{\gd}{\gd \xi_{\bga}^k\(z_1\)}\Langle \ww{Q}^{n}_{\bnu}\(z_2\)\Rangle=
\Langle  {\mathcal T} \ww{Q}^{n}_{\bnu}\(z_2\)\ww{Q}^k_{\bga}\(z_1\) \Rangle\\-\Langle  \ww{Q}^{n}_{\bnu}\(z_2\)\Ra\La\ww{Q}^k_{\bga}\(z_1\) \Rangle.
\end{multline}
If now we put together all components of \eq{eq:F.1} we get
\begin{widetext}
\begin{multline}
\frac{i\gd}{\gd \xi_{\bga}^k\(z_1\)} D^{m-k,n}_{\bgb,\bnu}\(z_1,z_2\)=
\Langle {\mathcal T}\ww{Q}^{m}_{\bgb\otimes\bga}\(z_1\)\ww{Q}^n_{\bnu}\(z_2\) \Rangle
-\Langle \ww{Q}^{m}_{\bgb\otimes\bga}\(z_1\) \Rangle \Langle \ww{Q}^{n}_{\bnu}\(z_2\)\Rangle\\
-\Langle {\mathcal T}\ww{Q}^{m-k}_{\bgb}\(z_1\)\ww{Q}^n_{\bnu}\(z_2\)\Ra\La\ww{Q}^k_{\bga}\(z_1\) \Rangle
+\Langle  \ww{Q}^{m-k}_{\bgb}\(z_1\)\Ra\La\ww{Q}^k_{\bga}\(z_1\) \Rangle \Langle \ww{Q}^{n}_{\bnu}\(z_2\)\Rangle\\
-\Langle  {\mathcal T} \ww{Q}^{n}_{\bnu}\(z_2\)\ww{Q}^k_{\bga}\(z_1\) \Rangle \Langle \ww{Q}^{m-k}_{\bgb}\(z_1\)\Rangle
+\Langle  \ww{Q}^{n}_{\bnu}\(z_2\)\Ra\La\ww{Q}^k_{\bga}\(z_1\) \Ra \Langle \ww{Q}^{m-k}_{\bgb}\(z_1\)\Rangle.
\label{eq:F.2}
\end{multline}
\eq{eq:F.2} finally gives
\begin{align}
\frac{i\gd}{\gd \xi_{\bga}^k\(z_1\)} D^{m-k,n}_{\bgb,\bnu}\(z_1,z_2\)= D^{m,n}_{\bgb\otimes\bga,\bnu}\(z_1,z_2\)
-\La \ww{Q}^k_{\bga}\(z_1\) \Ra D^{m-k,n}_{\bgb,\bnu}\(z_1,z_2\)
-\La \ww{Q}^{m-k}_{\bgb}\(z_1\)\Ra D^{k,n}_{\bga,\bnu}\(z_1,z_2\).
\label{eq:F.3}
\end{align}
\end{widetext}
In \eq{eq:F.2} we have used the fact that
\begin{align}
\Langle {\mathcal T}\ww{Q}^{m-k}_{\bgb}\(z_1\)\ww{Q}^n_{\bnu}\(z_2\)\ww{Q}^k_{\bga}\(z_1\) \Rangle=\Langle {\mathcal
T}\ww{Q}^{m}_{\bgb\otimes\bga}\(z_1\)\ww{Q}^n_{\bnu}\(z_2\) \Rangle.
\end{align}
Eq.\eqref{eq:F.3} proves \eq{eq:D_as_dD}. 

\section{Summary of definitions}\label{definitions}
\paragraph{Bosonic coordinates and the interaction vertex}
\begin{gather*}
\ww{Q}^n_{\bnu}=\prod_{i=1}^n \ww{Q}_{\nu_i}.\\
V^n_{\bnu}\(\xx\)=\frac{1}{n!}\(\prod_{i=1}^n \partial_{\nu_i}\)_{eq} V_\text{e--b}\(\xx\).\\
\hat \gc^{n}_{\bnu} = \int\!d\xx\,\hat \psi^\dagger\(\xx\)V^{n}_{\bnu}\(\xx\) \hat \psi\(\xx\).
\end{gather*}
\paragraph{Auxiliary fields}
\begin{align*}
  \hat{H}_{\xi,\eta}\(z\)=\hat{H} + \sum_{n,\,\bnu}\xi^{n}_{\bnu}\(z\)\ww{Q}_{\bnu}^n
  + \int\! d\xx\, \eta\(\xx,z\)\hat{\gr}\(\xx\).
\end{align*}
\paragraph{Correlators and electronic response}
\begin{gather*}
  G(1,2)\equiv -i\Langle\mathcal{T}\{\hat \psi\(1\)\hat\psi^\dagger\(2\)\}\Rangle.\\
  D^{n,m}_{\bmu,\bnu}\(z_1,z_2\)\equiv
  -i\Langle\mathcal{T}\wl\Delta\ww{Q}^n_{\bmu}\(z_1\)\Delta\ww{Q}^m_{\bnu}\(z_2\)\wr\Rangle.\\
  \chi\(1,2\) \equiv \frac{\delta \langle \hat \gr\(1\)\rangle}{\delta \eta(2)}.
\end{gather*}
\paragraph{Mean--field potentials}
\begin{gather*}
  \Phi(1)=\sum_{m,\,\bmu}V_{\bmu}^{m}\(\xx_1\)\Langle \ww{Q}_{\bmu}^m\(z_1\)\Rangle.\\
   U_{\mu,\nu}\(z_1\)=\sum_{n,\,\bku} n\,\langle\hat\gamma^n_{\mu\oplus\bku\oplus\nu}\(z_1\)\rangle \Langle\ww{Q}^{n-2}_{\bku}\(z_1\)\Rangle.
\end{gather*}
\paragraph{Electronic mass operator}
\begin{gather*}
  M\(1,2\)=i\sum_{n,\,\bnu}\sum_{m,\, \bmu}\int\!\!d3\!\int\!dz_4\,V_{\bnu}^{n}\(\xx_1\) G(1,3)\\
  \quad\times \overline{\Gamma}_{\bmu}^{\text{e-b},m}\(3,2;z_4\) D^{m,n}_{\bmu,\bnu}\(z_4,z_1\).
\end{gather*}
\paragraph{Bosonic mass operator}
\begin{multline*}
  \Pi_{\mu,\nu}\(z_1,z_2\)=\sum_{I=\(1,2a,3,4a,4b,4c,4d\)}\Pi^{\(I\)}_{\mu,\nu}\(z_1,z_2\)\\+\Pi^{\(2b\)}_{\mu,\nu}\(z_1\)\gd\(z_1-z_2\).
\end{multline*}
\paragraph{Vertex functions}
\begin{gather*}
\Gamma^{\text{e-e}}\(1,2;3\)= \frac{\delta G^{-1}\(1,2\)}{\delta\eta\(3\)}.\\
\Gamma_{\bku}^{\text{e-b},k}\(1,2;z_3\)= \frac{\delta G^{-1}\(1,2\)}{\delta\xi_{\bku}^k\(3\)}.\\
\overline{\Gamma}^{\text{e-b},k}_{\bku}\(1,2;z_3\)= \frac{\delta G^{-1}\(1,2\)}{\delta \Langle Q_{\bku}^m\(z_3\)\Rangle}.\\
\Gamma^{\text{b-e}}_{\mu,\nu}\(z_1,z_2;3\)=\frac{\delta D^{-1}_{\mu,\nu}\(z_1,z_2\)}{\delta\eta\(3\)}.\\
\Gamma^{\text{b-b},k}_{\mu,\nu;\bku}\(z_1,z_2;z_3\)=\frac{\delta D^{-1}_{\mu,\nu}\(z_1,z_2\)}{\delta\xi_{\bku}^k\(3\)}.
\end{gather*}
\paragraph{Kernels}
\begin{gather*}
  K^\text{e-e}\(1,5;2,4\)=\frac{\delta \Sigma^{\it e}\(1,2\)}{\delta G\(4,5\)}.\\
  K^\text{e-b}\(1,z_5;z_2,4\)=\frac{\delta M\(1,2\)}{\delta D_{\phi,\psi}\(z_4,z_5\)}.\\
  K^\text{b-e}\(z_1,5;2,z_4\)=\frac{\delta \Pi_{\mu,\nu}\(z_1,z_2\)}{\delta G\(4,5\)}.\\
  K^\text{b-b}\(z_1,z_5;z_2,z_4\)=\frac{\delta \Sigma^{\it b}_{\mu,\nu}\(z_1,z_2\)}{\delta D_{\phi,\psi}\(z_4,z_5\)}.
\end{gather*}

\bibliography{paper}
\end{document}